\documentclass[num-refs]{wiley-article}

\usepackage{siunitx}

\papertype{Original Article}
\paperfield{Journal Section}

\title{Flood risk map from hydrological and mobility data: a case study in São Paulo (Brazil)}

\author[1\authfn{1}]{Lívia Rodrigues Tomás}
\author[2]{Giovanni Guarnieri Soares}
\author[2]{Aurelienne A. S. Jorge}
\author[3]{Jeferson Feitosa Mendes}
\author[4]{Vander L. S. Freitas}
\author[1,2]{Leonardo B. L. Santos}

\contrib[\authfn{1}]{Equally contributing authors.}

\affil[1]{National Center for Monitoring and Early Warning of Natural Disasters (Cemaden), São José dos Campos, São Paulo, 12245-230, Brazil}
\affil[2]{National Institute for Space Research (INPE), São José dos Campos, São Paulo, 12227-010, Brazil}
\affil[3]{Department of Environmental Engineering, São Paulo State University, São Paulo, 12245-00, Brazil}
\affil[4]{Department of Computing, Federal University of Ouro Preto, Ouro Preto, Minas Gerais, 35400-000, Brazil}

\corraddress{Lívia Rodrigues Tomás, National Center for Monitoring and Early Warning of Natural Disasters (Cemaden), Brazil}
\corremail{liviatomas@gmail.com}

\fundinginfo{FAPESP, Grant Number: 2015/50122-0, 2018/06205-7; DFG-IRTG, Grant Number: 1740/2; CNPq, Grant Number: 420338/2018-7}

\begin{document}

\begin{frontmatter}
\maketitle

\begin{abstract}
ABSTRACT

\noindent Cities increasingly face flood risk primarily due to extensive changes of the natural land cover to built-up areas with impervious surfaces. In urban areas, flood impacts come mainly from road interruption. This paper proposes an urban flood risk map from hydrological and mobility data, considering the megacity of São Paulo, Brazil, as a case study. We estimate the flood susceptibility through the Height Above the Nearest Drainage algorithm; and the potential impact through the exposure and vulnerability components. We aggregate all variables into a regular grid and then classify the cells of each component into three classes: Moderate, High, and Very High. All components, except the flood susceptibility, have few cells in the Very High class. The flood susceptibility component reflects the presence of watercourses, and it has a strong influence on the location of those cells classified as Very High.

\end{abstract}
\end{frontmatter}

\section{Introduction}

Cities increasingly face a variety of hazards primarily due to extensive changes of the natural land cover to built-up areas with impervious surfaces, along with climate change and inadequate management \cite{LI2018, YIN2015, WU2012, OGDEN2011}. The potential for a hazard to become a disaster depends on the degree of exposure of a population and its physical or economic assets, coupled with their respective vulnerability \cite{DICKSON2012}.

Floods have severely impacted many cities, being one of the most frequent and damaging natural hazards worldwide \cite{QUIROS2020}. Floods affect millions of people around the globe and result in significant disruption to the built environment in a community \cite{Jongman2015a, Nofal2021}. In Brazil, the sudden and gradual floods correspond to 50\% of the occurrences of disasters recorded in recent years \cite{UFSC2013}.

In the municipality of São Paulo, Brazil, the largest city in Latin America, almost 12.4 million people live in 1,521 km² \cite{IBGE2020}, and there are 8.7 million vehicles registered, which corresponds to 8.17\% of the total number of vehicles in Brazil \cite{DENATRAN}. On average, a São Paulo inhabitant spends 2 hours and 25 minutes with daily commutes \cite{IBOPE}. Vale \cite{vale2020} says that opportunity costs arising from urban immobility can reach 1.4 billion dollars per year in São Paulo.

This megacity is experiencing some consequences of climate change: more frequent heavy rains and floods, higher temperatures, and decreased air humidity \cite{DICKSON2012}. In addition to economic losses, damage caused by floods impairs the mobility of people who live in or pass through flooded areas. In São Paulo, several factors related to the hydrographic basin, local catchments, topography, and land use and occupation, for instance, cause floods \cite{Morales2020, FELIPE}.

According to Traffic Engineering Company's (CET) data, the city averaged 77 and 85 kilometers of traffic jams in the morning and afternoon peak hours, respectively, in 2019. Massive traffic events can result from the increased number of vehicles, unplanned urbanization, accidents on the roads, heavy precipitations, and flooding. Heavy precipitations usually result in slower traffic and more dangerous traffic conditions since it causes bad visibility and changes road friction \cite{litzinger, FELIPE}.

For most cities in developing countries, collecting reliable and timely data is a challenging task, especially in the construction of risk scenarios from a multidisciplinary perspective. Integrating knowledge from multiple fields better reflects the geographic environment in which various natural processes, human activities, and information interactions exist. In this sense, methodologies comprised of hydrology, mobility, and Geographic Information Systems (GIS) have proved effective in many case studies to model complex urban systems  \cite{LI2018}. Furthermore, GIS are of great help to analyze data and produce meaningful information for urban management and resilience planning.

This paper presents an urban flood risk map combining elements from natural geographies, such as hydrological indexes, and elements from human geographies, such as urban mobility data, in both cases using different sources and geographic units. To estimate the exposure ($E$) component, we consider the resident population and the population that works and studies within the study area. The vulnerability ($V$) component takes into account the local vulnerability ($LV$) and the network vulnerability ($NV$) of the road network. The flood susceptibility ($FS$) is estimated by the HAND (Height Above the Nearest Drainage) algorithm.

\section {Theoretical Background} \label{section:LitRev}

We can use the relationship between hazard and vulnerability to discuss risk scenarios. Hazard refers to a potentially harmful natural process or phenomenon occurring in a given location and within a specified period. Vulnerability is the set of processes and conditions resulting from physical, social, economic, and environmental factors, which increases the susceptibility of a community (element at risk) to the impact of hazards. Vulnerability comprises both physical aspects (resistance of buildings and infrastructure protections) and human factors, such as economic, social, political, technical, cultural, educational, and institutional \cite{UNISDR2009, WISNER2004}.

The United Nations Office for Disaster Risk Reduction \cite{UNISDR2009} defines disaster risk as to the potential loss of life, injury, or destroyed or damaged assets that could occur to a system, society, or a community in a specific period, determined stochastically as a function of hazard, exposure, vulnerability, and capacity.

Other variables can be included in risk analysis such as number of deaths, number of people affected as a result of a disaster, demographic density (inhabitant/km²), poverty index, elderly population, municipal human development index, number of events per year, total resident population and municipality area \cite{UNDP2004}.

According to Kron \cite{KRON2005}, the scientific community widely agrees that risk is the product of a hazard and its consequences. There is no risk in a region where there are no people or values that can be affected by a natural phenomenon. Therefore, the risk is always present in urbanized areas with people, constructions, and road infrastructure. Three components determine the risk \cite{CRICHTON2007}: 

\begin{itemize}
    \item the hazard: the threatening natural event including its probability of occurrence;
    \item the exposure: the values/humans that are present at the location involved;
    \item the vulnerability: the lack of resistance to damaging/destructive forces.
\end{itemize}

In the case of the hazard being flooding, this is a hydrological phenomenon caused by an excess capacity of surface runoff and urban drainage systems forming accumulations of water in impermeable areas \cite{CENAD2014}, it is usually associated with rain. According to Walesh \cite{WALESH1989}, there are two main types of floods caused by rainfall. The first is a large amount of rainfall, occurring at a relatively low intensity over a long period and on a large area. The second is caused by high-intensity thunderstorms occurring over small regions.

As for Kron \cite{KRON2005}, the three main types of flooding are: storm surge, river flood, and flash flood. Flash floods occur when water quickly sweeps over an area which is challenging to deal with, and it is not easy to predict the amount of rain expected within the spatial area over a short period \cite{GLAGO2021}.

When floods occur in urban areas during heavy rainfall, causing momentary accumulation of rainwater in certain places due to a deficiency in the drainage system, it is called Surface Water Flooding (SWF). It includes pluvial flooding, sewer flooding, flooding from small open-channel and culverted urban watercourses, and overland flows from groundwater springs. The predominant cause of SWF is short-duration intense rainfall, occurring locally \cite{Falconer2009}.

The impacts of floods occur at different spatiotemporal scales: from intra-urban mobility to inter-urban, in a period that can vary from hours to months or even years, in some cases \cite{GLAGO2021, LONDE2017}. Regarding SWF, the impacts on daily activities come mainly from infrastructure disruption, such as inundated roads that block people's daily routes. People often stand up to high risk when commuting in bad weather since they cannot cancel regular trips \cite{LIU2021}.

The characteristics of urban drainage hinder the management of SWF. Land use has a significant influence on surface water behavior, wherein the presence of built-up areas raises the volume of surface water runoff \cite{KAZMIERCZAK2011, GILL2007}.

Several researchers have analyzed the interactions between elements of a risk framework in recent years. Beden and Keskin \cite{Beden2021} produced a flood map and evaluated the flood risks in case of insufficient flow data in Turkey; Kazmierczak and Cavan. \cite{KAZMIERCZAK2011} estimated the SWF risk, encompassing hazard, vulnerability, and exposure, to urban communities in Greater Manchester, UK. Ramkar and Yadav \cite{Ramkar2021} created a flood risk index in data-scarce river basins using the Analytical Hierarchical Process and GIS approach. Zokagoa et al. \cite{Zokagoa2021} mapped the flood risk using uncertainty propagation analysis on a peak discharge in Quebec.

Beden and Keskin \cite{Beden2021} produced a hydraulic model and flood map of the Ceviz Stream in Turkey, where many people live and work, and there are a large number of private properties and public infrastructure. The authors used the MIKE FLOOD software to develop the hydraulic model. The modeling phase generated water depth, velocity, and flood extent area data used as parameters in flood risk assessment. 

Kazmierczak and Cavan \cite{KAZMIERCZAK2011} explored the spatial distribution of SWF, the vulnerability of communities to flooding, and the characteristics of the physical environment and land use that affect people's exposure to flooding. They used four indicators for the vulnerability of people to flooding, and the analysis of the presence and spatial distribution of SWF areas, land use types, green cover, and housing to perform a spatial association between hazard, vulnerability, and exposure. Their results indicate that some of the most vulnerable people are at high risk of flooding due to socioeconomic characteristics of the population, spatial distribution of the hazard, and the land use and housing types present in the area.

Ramkar and Yadav \cite{Ramkar2021} developed a flood risk index map using an integrated approach of Geospatial technique and Multiple Criteria Decision-Making Technique. The flood risk index was calculated by integrating the flood hazard and vulnerability maps. The flood hazard map considered slope, distance from the main river, land use, land cover, soil, drainage density, and rainfall. The vulnerability index took into account population density, crop production, and density of road-river intersection. The flood risk map results from the multiplication of flood hazard and vulnerability maps.

Zokagoa et al. \cite{Zokagoa2021} produced a probabilistic flood map using uncertainty propagation in real flood events in Quebec. They adopted a Monte Carlo method to generate an ensemble of results from different combinations of uncertain input parameters. Then, the weighted average of ensemble results is used to derive a probabilistic floodplain map.

Other authors estimated one component of the risk of flooding, as Liu et al. \cite{LIU2021} that estimated the exposure; and Praharaj et al. \cite{Praharaj2021}, Kasmalkar et al. \cite{Kasmalkar2020}, and Lu et al. \cite{LU2015} which estimated the impact of flooding in transport infrastructure.

Liu et al. \cite{LIU2021} proposed an approach to estimate flood-affected populations by combining mobility patterns with multi-source data. They used the Gravity model and Monte Carlo simulation, together with points of interest and building footprint data, to model automobile commute patterns. The locations of impassable roads were retrieved by social media data with real-time inundation records and flood hazard maps with inundation depths. Finally, they estimated the affected population employing the difference in commute time between no-flood and flood conditions.

Praharaj et al. \cite{Praharaj2021} quantified transportation impacts of recurring flooding using a predictive model. Traffic volumes and flood incidents were estimated through a combination of agency-provided and crowdsourced data. Hydrological data include rain and tidal gauge data. The authors concluded that the impact of recurring flooding events on transportation networks is local; thus, they do not recommend a citywide or regional analysis due to the heterogeneous effects of flooding across various links.

Kasmalkar et al. \cite{Kasmalkar2020} integrated a traffic model with flood maps to simulate regional traffic patterns in the San Francisco Bay Area in the presence of coastal flooding. They used a road network model, origin-destination commuter data for weekday morning commutes, number of employees who reside and work in a given census block, and flood maps. Their model scale is associated with census blocks that range from 100 m to 10 km in length. Traffic analyses were conducted on zones that range from 1 to 50 km in length. Their model highlights traffic flow disruption caused by flooding.

Lu et al. \cite{LU2015} proposed a road prioritization methodology based on a location-based accessibility index. This index measures the network-wide performance before and after transportation network interdiction and quantifies the degree of network degradation. The methodology is applied to the road network threatened by flood risk from storm surge, sea-level rise, and intense precipitation. The results show that some infrastructure is critical to adjacent areas, while some become important to a broader region.

The references cited in this section demonstrate the diversity of existing methods for estimating flood risk and its associated components. There are stochastic and deterministic methods, and others use pre-existing data from one or more components and those that estimate all components. 

There is no one method better than another for all cases, but one that is better adapted to the characteristics of the study area. Building upon the literature review, we present the material and methods adopted in this work.

\section{Material and Methods} \label{section:MM}

\subsection{Study Area and Datasets} \label{section:StudyArea}

\begin{figure}[htb!]
\centering
\includegraphics[width=10cm]{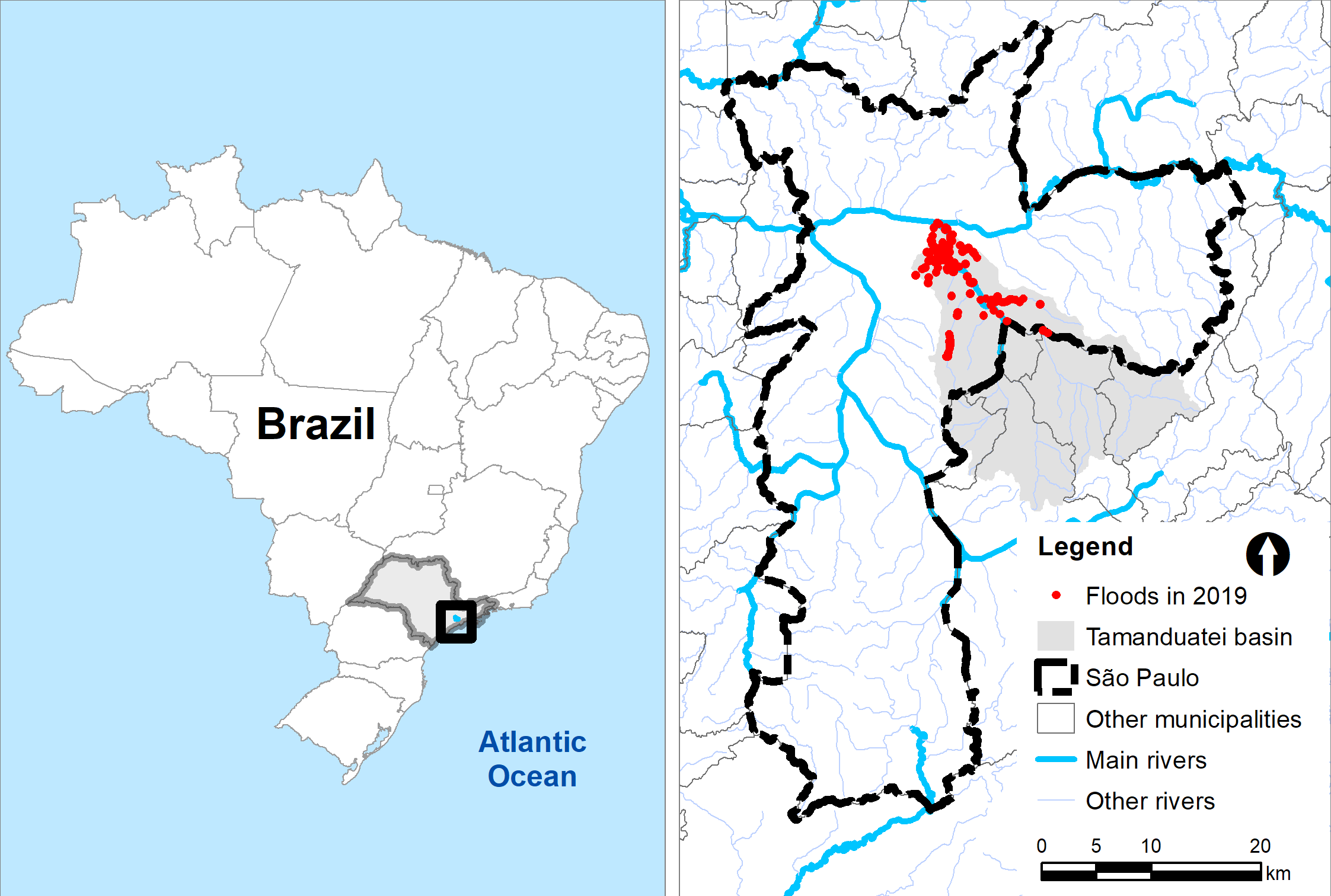}
\caption{Location of the study area in the Tamanduateí basin with flood occurrences in 2019.}
\label{Fig_StudyArea}
\end{figure}

\begin{figure}[htb!]
\centering
\includegraphics[width=12cm]{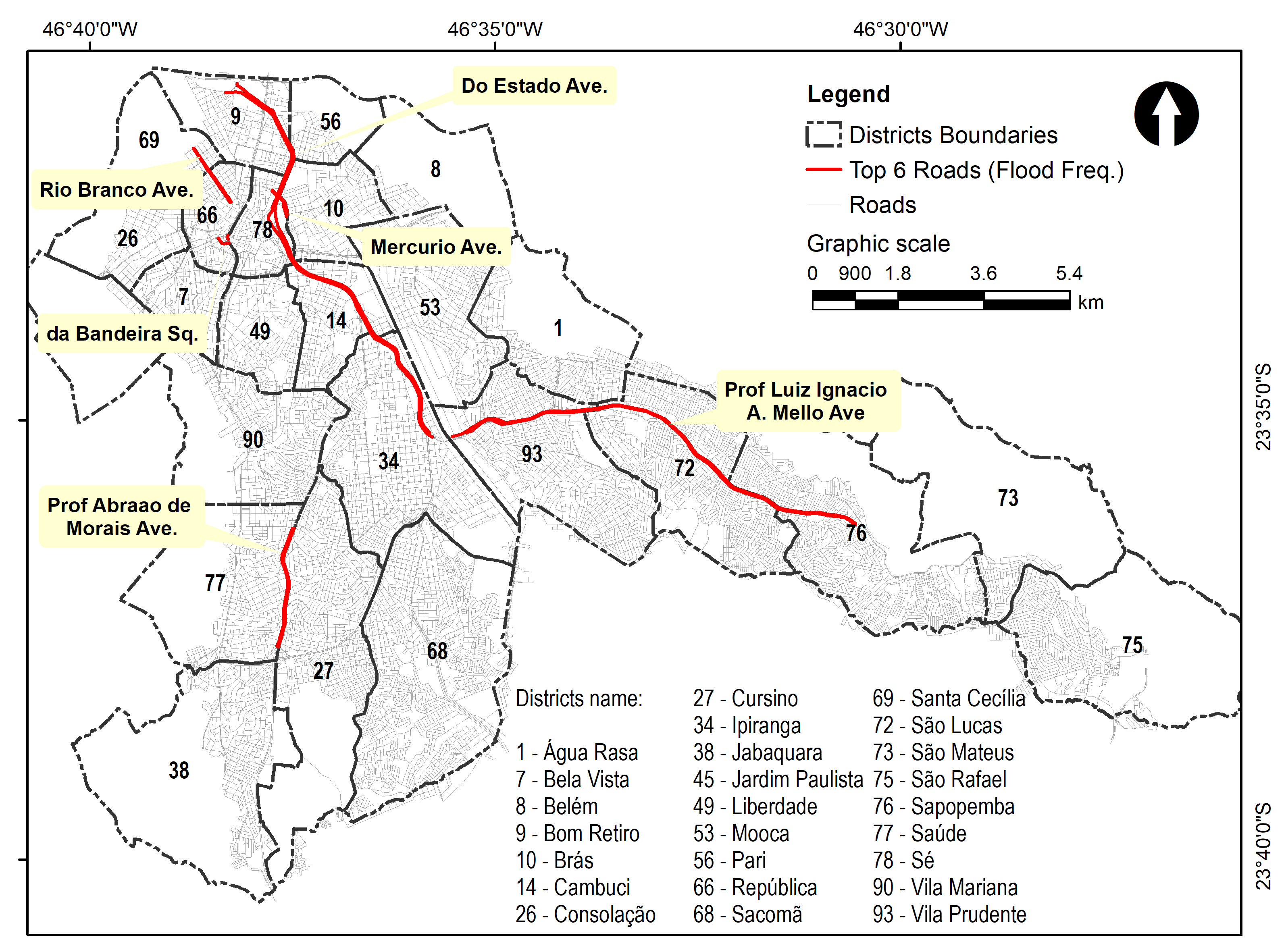}
\caption{Study area districts and the top 6 roads in flood frequency in 2019.}
\label{Fig_Districts}
\end{figure}

The study area comprises 115.3 km² of the Tamanduateí river basin within the municipality of São Paulo, the capital of the state of the same name, and the leading financial center of Brazil (Figure \ref{Fig_StudyArea}). More than 80\% of the Tamanduateí riverbed is impermeable, and floods along the riverbanks are constant \cite{PEREIRA2006}. This basin presents a high number of extreme rainfall events in the city \cite{COELHO2016}. Figure \ref{Fig_StudyArea} also shows the occurrences of flooding with road disruptions in the Tamanduateí basin in 2019. The roads with the highest frequency of flooding in 2019 are highlighted in Figure \ref{Fig_Districts}.

In this hierarchical order, the city of São Paulo has some official administrative divisions: municipality, sub-prefectures, and districts. The study area covers 25 districts (some of them, partially), as shown in Figure \ref{Fig_Districts}. The Sé and Repúlica districts form the historical center, the original nucleus of the city, where approximately 600 thousand people circulate per day \cite{PSP2021}. Some famous places in the city, such as The Sé Square, The Anhangabaú Valley, and the Municipal Market, are located in the Sé District (Figure \ref{fig:zoom}). Although they are not a unit of analysis, districts are used throughout the text and maps to highlight relevant locations in the results.

We use data from different sources and geographic topologies. Therefore we chose the Brazilian statistical grid \cite{IBGE2016} to integrate and analyze data. Geographically and socially speaking, this unit is arbitrary, not having a meaning that can be transported to the real world, as it does not consider the distribution of any underlying process or phenomenon \cite{GRASLAND2006}. However, the cells perfectly serve the purpose of a receptacle, remaining stable over time, presenting a regular and simple shape, with small enough dimensions to act as bricks in the construction of any desired geographic outline \cite{NORMAN2003, GUZMAN2013}, in addition to meeting the demands of data dissemination for small areas and integrate incompatible geometries or different administrative boundaries.

The Brazilian statistical grid \cite{IBGE2016} has cells with dimensions of 1 km x 1 km in rural areas and 200 m x 200 m in urban areas. Except for one cell, the entire study area is urban. In this work, we use the attribute ``resident population'' by cell and the geometry of the grid. We use two vector data to select the cells from the grid that form the study area: São Paulo municipality boundary and Tamanduateí river Basin Boundary. Their intersection gives 3,087 cells from which we exploit the following data:
\begin{itemize}
    \item Resident population by cell from 2010 Brazilian census \cite{IBGE2016};
    \item Geographic coordinates of Workplaces and the number of people who work at each location from Origin Destination Research (OD Database) \cite{METROSP2017};
    \item Geographic coordinates of Educational institutions and the number of people who study at each location from OD Database \cite{METROSP2017};
    \item Flood registries in 2019 obtained from Emergency Management Center of the city of São Paulo - CGE website \cite{CGE2019};
    \item Road system vector data from Center for Metropolitan Studies - CEM \cite{CEM2021};
    \item SRTM Digital Elevation Model - DEM \cite{Farr2007}.
\end{itemize}

In the first step of the methodology, we perform an exploratory analysis to better understand the characteristics of the datasets and the relationships between the analyzed variables and then define the best conceptual model. Thereafter, we estimate the $FS$ and potential impact ($PI$) components to calculate the urban flood risk ($R$). The $PI$ comes from the $E$ and $V$ components.

\subsection{Flood Susceptibility}\label{section:FS}

Topography is a hydrologic driver, defining the speed and direction of flows, which defines hydrological relations between different points within a basin. Considering a DEM represented by a grid, the simplest and most widely used method to determine flow directions is the D8 (eight flow directions) \cite{OCallaghan1984}. The flow goes to the steepest downward slope within the set of eight possible directions around a grid point. Based on this approach, it is possible to determine a synthetic drainage network under a drainage threshold - a cell is considered in the drainage network if the upstream area associated with it is equal or greater than the threshold. Details in \cite{RENNO2008, NOBRE2011}.

Assuming that all points belong to a flow path and that all flow paths are associated with respective drainage points, it is possible to define the Height Above the Nearest Drainage (HAND) of any given point. Therefore, the HAND map is a robust and versatile digital terrain model normalized by the drainage network. Low HAND's values represent points with altimetry close to the altimetry of the nearest drainage element. Thus the lower the HAND, the more significant the flood probability of occurrence \cite{RENNO2008, NOBRE2011}. We use this well-established concept to create the classes of flood susceptibility.

We use the SRTM DEM model \cite{Farr2007} as input in the TerraHidro plugin, in the TerraView software \cite{TerraView}, with a drainage threshold of $5000$ cells to obtain the HAND raster with a spatial resolution of 30 meters. Since the HAND raster has different geometry and spatial resolution from the grid we use to aggregate the variables, each grid cell receives the minimum HAND value among those that intersect it. The HAND values (in meters) aggregated in the grid cells are then classified, as detailed in section \ref{section:Classes}, to obtain the $FS$ classes.

\subsection{Exposure}\label{section:Exposure}

Damage caused by disasters impairs the mobility and accessibility of people living in or moving through the affected areas. Considering these two situations, we use the resident population, initially aggregated by grid cell, and the number of people who work or study in the study area (obtained from OD Database), disregarding people who work from home, to represent people moving through the study area. People's main reasons for moving are work and study, justifying our choice.

The São Paulo Subway Company conducted the OD Research in 2017, which encompasses all 39 municipalities of the metropolitan region. The OD Database is available on the São Paulo Subway Company website \cite{METROSP2017}. The geographic coordinates of workplaces and study institutions and the number of people who commutes to each point are available in the database. Using these fields, we create a new vector data containing points representing workplaces and study institutions with the attribute ``number of people that commutes to each point''.

The commuting pattern results in a particular picture of exposure, with variation in space and time. The consequences of a disaster vary in size, if it occurs on a typical workday, a weekend, or holiday, for example. On a typical working day, roads have higher traffic, streets are much more crowded, and people are at schools or work instead of being home. Thus, people's exposure is directly related to the time a disaster occurs, influencing the degree of exposure. Legeard, as cited in Veyret \cite{VEYRET2007}, argued that, once this dynamic nature of mobility processes is well known, it is possible to produce exposure maps by time slots:

\begin{itemize}
\item Daily exposure: Commercial period, except for peak hours.
\item Peak exposure: Period of active movement in transportation net (roads, collective transportation, stations).
\item Night exposure: When the population is concentrated in residential areas.
\end{itemize}

Then, we aggregate the number of commuters, from a point vector data, in the regular grid. Each grid cell has the sum of commuters and resident population, resulting in the exposed population.

\subsection{Vulnerability}\label{section:Vulnerability}

Data of extreme rainfall events in the year 2019, obtained from the CGE, was used to estimate $LV$. We chose 2019 because it was the year before the beginning of mobility restrictions due to SARS-CoV-2 in Brazil.

The start and end time of road blocking and the address (street name and intersection) where there is a flood is available on the CGE website \cite{CGE2019}. We tabulate this information and extract the Universal Transverse Mercator - UTM coordinates for each point, using vector data of the São Paulo municipality's street network. Then, we create vector data with 1,165 points representing traffic blocks due to floods. Ultimately, we intersect the points with the regular grid to select occurrences in the study area.

\begin{figure} [!htb]
    \centering
    \includegraphics[scale=0.3]{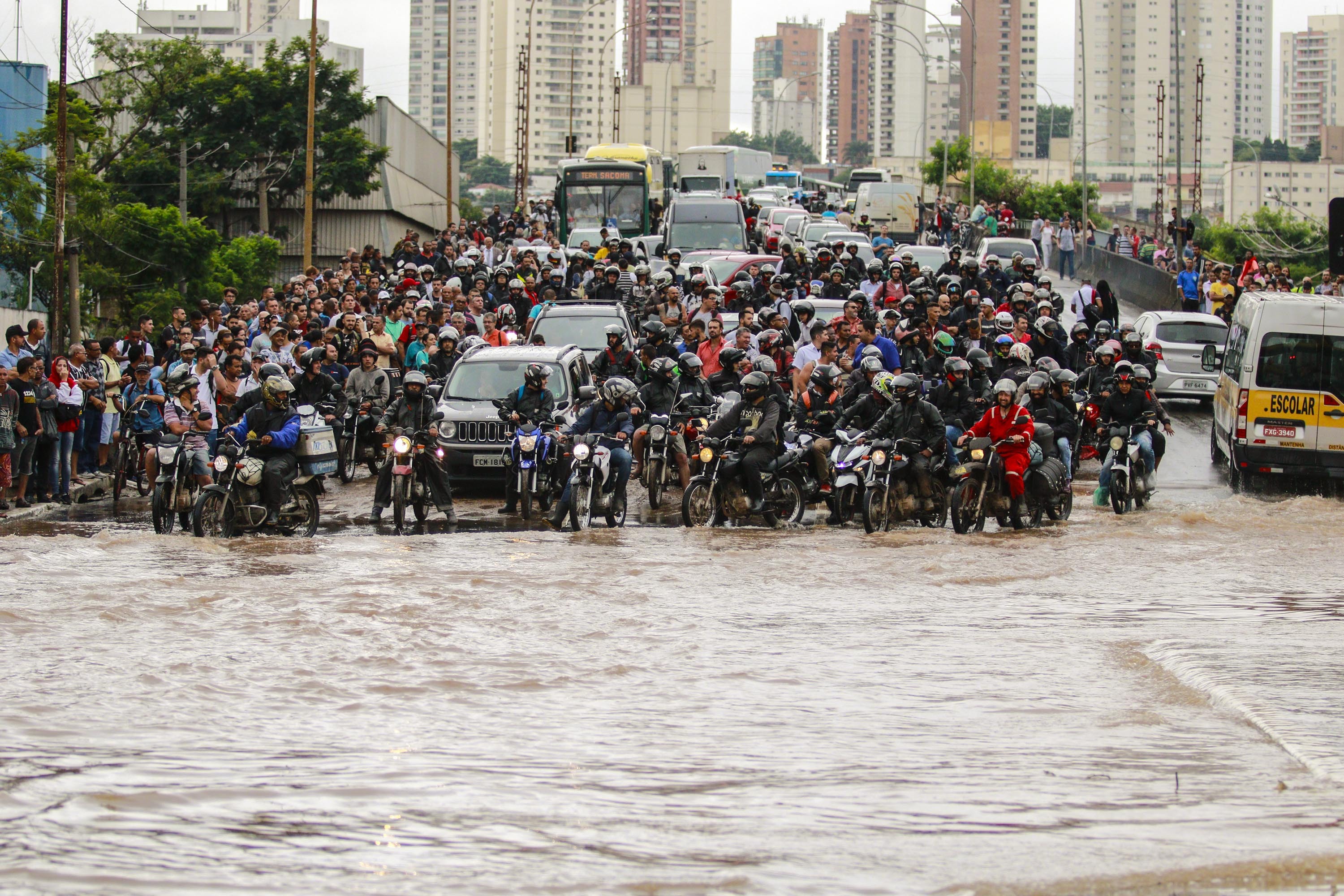} \\
    \caption{Street block caused by an extreme rainfall event in the Tamanduatei River in March 2019 \cite{fotoalag}.} 
    \label{fig:roadblock}
\end{figure}

There were 1,165 traffic blocks due to floods in São Paulo municipality in the year 2019 \cite{CGE2019}, of which 280 were inside the study area. These points correspond to locations where streets needed to be blocked due to extreme rainfall events. We sum up the street block duration in each cell to find the $LV$. Figure \ref{fig:roadblock} illustrates a street block caused by an extreme rainfall event in the Tamanduateí River, in the city of São Paulo, in March 2019.

The second variable used to calculate $V$ is the $NV$, which depends on the road system vector data \cite{CEM2021}, which has 23,702 lines and represents the roads. In addition to geometry, this file has alphanumeric attributes, such as the road hierarchy (or road function classification) that we use in the exploratory analysis of Section 4.1.

Roads serve as an essential means for people to travel from one place to another, connecting regions and neighborhoods and accessing properties. The characteristics of the road system can be investigated through a complex network, a structure in which connections (edges) link pairs of elements (nodes), forming a graph with particular topological properties \cite{Santos2019Springer}.

We use the Gis4Graph tool \cite{GeoINFO2017} to build a geographical graph from the road system. The adopted methodology is based on the concept of (geo)graph, a graph whose vertices are attached to a geographical location and edges with an intrinsic spatial dependence \cite{Santos2017}. The process consists of identifying the intersections between the lines. As a result, each road becomes a node, and each intersection is an edge in our graph. Furthermore, the tool produces vector data with the resulting network to make it possible to analyze it in a GIS environment \cite{GeoINFO2017}. The (geo)graph's approach is applied in a variety of different domains \cite{Santos2017, Santos2019Springer, TTC, Ceron, Lamosa2021}.

The role of a road system is to provide adequate fluidity of the travel demand. The operation of the road system is subject to hazards that could negatively affect the level of service. The hazards can be related to meteorological events that affect vehicular traffic, for example. The analysis of vulnerability in a road system can be understood as an assessment of the behavior of the road network when it suffers interference from adverse events (unexpected or undesirable) in the elements that comprise it \cite{MURRAY2007}.

In this study, we estimate the topological vulnerability of the (geo)graph to understand how the network behaves in the absence of a road for whatever reason, using the \textit{igraph} module for Python \cite{igraph}. The method consists in evaluating performance changes in the network when disconnecting nodes from the grid \cite{Latora2005}. We calculate the $NV$ to find important links and components in the (geo)graph. We use efficiency as performance, and the inverse of the shortest path between nodes as efficiency \cite{Goldshtein2004}. Thus, we start by calculating the network's global efficiency to use it as a performance. Then, we go node by node, disconnect it and recalculate the new global efficiency for each case. We calculate the node vulnerability for each node, independently of what happens in other nodes, as follows

\begin{equation}
    nV(i) = \frac{\mathcal{E}(G) - \mathcal{E}(G_{i})}{\mathcal{E}(G)},
\end{equation}

\noindent being $nV(i)$ the node vulnerability, $\mathcal{E}(G)$ the original global efficiency and the $\mathcal{E}(G_{i})$ the global efficiency of $G$ considering the disconnection of node $i$.

Since our network has about 20,000 nodes and 50,000 edges, calculating the shortest paths and removing each node is a resource and time-demanding task, which we solve via High-Performance Computing, using Python's multiprocessing module. The computing architecture we use consists of two AMD Ryzen Threadripper 3960X 24-Core Processor, 128GB of RAM, and an NVIDIA Titan XP.

Using this method, we obtain a line vector file with node vulnerability ranging between $0$ and $1$. A lower node vulnerability value leads to a less relevant node regarding the network robustness. The set of nodes vulnerabilities forms the $NV$. Each grid cell receives the maximum value of the node vulnerability of the intersected lines, considering their spatial location. The aggregation follows

\begin{equation}
    NV(i) = \max \{ nV(j), \; \; j \in C_i \},
\end{equation}

\noindent where $NV(i)$ is the network vulnerability, $nV(j)$ is the node vulnerability, $i$ is the cell index, and $C_i$ is the set of streets that intersect cell $i$.

The combination of the $LV$ and $NV$ to form the $V$ component is detailed in the next section.

\subsection{Component Classification }\label{section:Classes}

After we aggregate the $E$, $LV$, $NV$, and $FS$ components into grid cells, we classify the cells of each component into three classes: Moderate, High, and Very High. We use the Jenks Natural Breaks algorithm \cite{Jenks1971} to define the thresholds of the classes (except for the $LV$). They are based on natural groupings inherent in the data. The algorithm identifies breakpoints by minimizing the variances within each class. In this way, similar points are grouped. 

The distribution of the $LV$ is very skewed; 96\% or 2,971 of the cells have no flooding occurrence, and therefore have no duration. So, we classify all cells with a value less than one hour as Moderate. The remaining cells were manually divided, observing the frequency graph (Figure \ref{graf:localvulnerability}).

The table \ref{tab:ClassCrossing} shows the resulting class when combining two classes.

\begin{table}[!htb]\centering
\begin{tabular}{r|r|r|r}
{\textbf{Classes}} & {\textbf{Moderate}} & {\textbf{High}} & {\textbf{Very High}} \\ \hline
\textbf{Moderate}   & Moderate  & Moderate & High   \\
\textbf {High}   & Moderate  & High & Very High   \\
\textbf{Very High}   & High  & Very High & Very High \\
\end{tabular}
\caption{Resulting class combining two classes.}
\label{tab:ClassCrossing}
\end{table}

The city of São Paulo is highly urbanized and dense. The city has a strong attraction for travel from neighboring cities, and its transport network has high traffic of vehicles and people daily. According to the OD Survey \cite{METROSP2017}, the number of trips exceeds 25 million on a typical day. Given this scenario, we propose that risk classification begins at a level Moderate because it inspires attention and monitoring. 

First, we cross the $LV$ with $NV$ to find $V$ classes, according to Table \ref{tab:ClassCrossing}. Next, we combine $V$ with $E$ classes to find the $PI$. Finally, the $R$ comes from the crossing between $PI$ and $FS$. The results of each combination are presented in Section \ref{section:Results}.

\section{Results and Discussion}\label{section:Results}

\subsection{Exploratory Analysis} \label{section:ExploratoryAnalysis}

Here are the results of the exploratory analysis of the data before aggregation into the grid. These results guide the method decision and give us an overview of the variables. We analyze the HAND, population, floods, and network vulnerability index.

\begin{figure} [!h]
    \centering
    \includegraphics[scale=0.6]{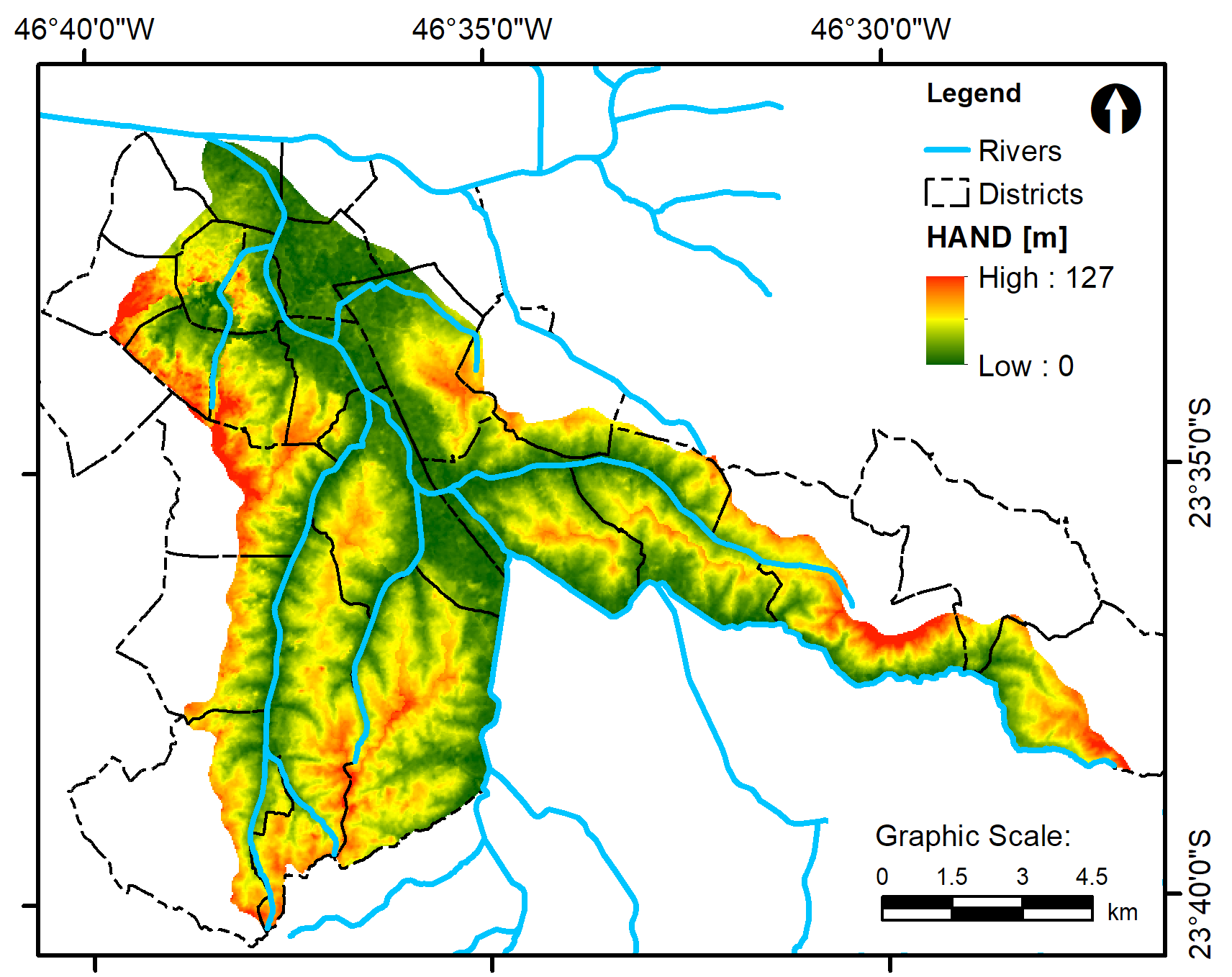} \\
    \caption{HAND raster in meters.}
    \label{fig:hand}
\end{figure}

The HAND raster has a spatial resolution of 30 m and 129,057 cells with values between 0 and 127 m (Figure \ref{fig:hand}). The average value of the cells is 29.40 m. Cells with a value equal to zero, those with the presence of watercourses, have the highest frequency with 3,310 cells or 2.6\%. Values between 0 and 10 m represent 25\% of cells. The figure \ref{graf:HAND} shows that smaller values are the most frequent. 

\begin{figure} [htb!]
    \centering
    \includegraphics[scale=0.33]{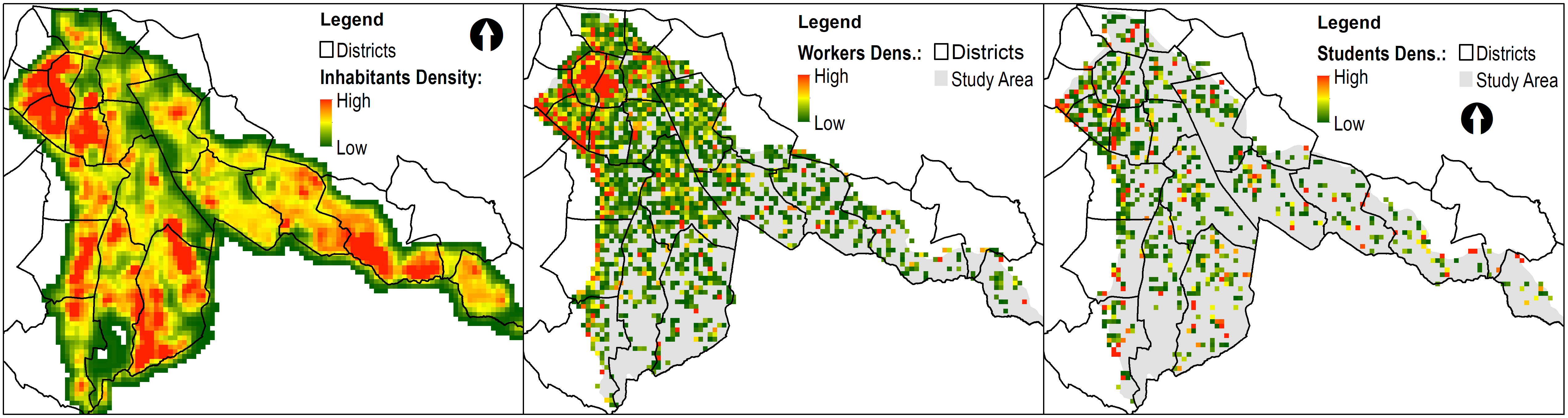} \\
    \caption{Kernel density of inhabitants, workers, and students.} 
    \label{fig:kernel}
\end{figure}

The resident population totals 1,637,265 and is spread out the study area, with only 7.7\% of cells (or 237 cells) without population. The cell with the highest value has 5,597 inhabitants, and the average value is 530. Considering the density of inhabitants, it is possible to notice some kernels, the most prominent being in Bela Vista, Consolação, and República districts (Figure \ref{fig:kernel}). Analyzing the OD survey data concerning workplaces, there are 4,560 points, of which 4,527 have different geographic coordinates. A total of 3,369,774 people work at these points. The place with the highest number of workers has 13,361 people, and the average value is 739 workers. Observing the density of workers, it is possible to notice two kernels, the biggest one in the historic center, in districts Sé and República (Figure \ref{fig:kernel}). Concerning study locations, there are 1,040 points, resulting in 1,390,969 students. Although it does not form a kernel, the concentration of students is more significant in five districts (number 7, 26, 49, 66, 78) (Figure \ref{fig:kernel}). The population that works and studies in the study area is 4,760,743 people, disregarding people who work from home.

\begin{figure}[!hbt]
    \centering
    \begin{minipage}{0.47\textwidth}
    \centering
    \includegraphics[scale=0.24]{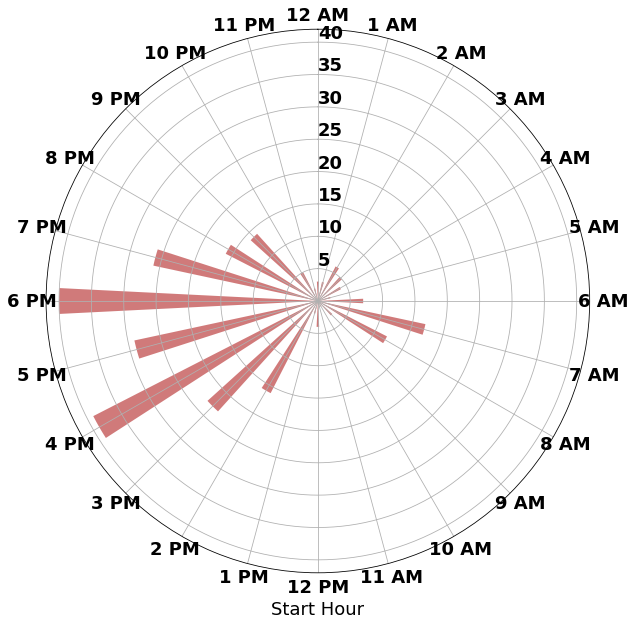} \\
    \caption{Number of floods per start time (in Hour).} 
    \label{fig:floodsstarttime}
    \end{minipage}
    \begin{minipage}{0.47\textwidth}
    \centering
    \includegraphics[scale=0.19]{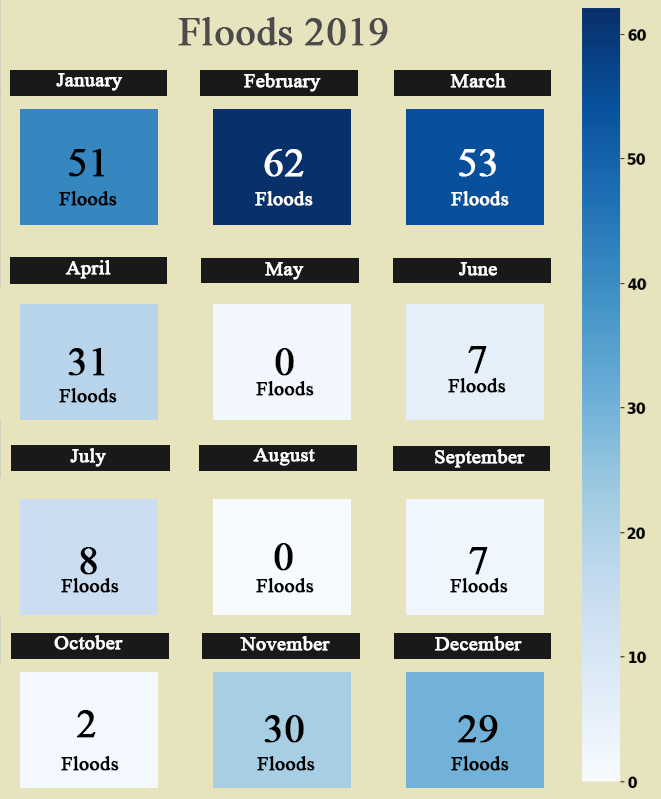} \\
    \caption{Number of floods per month.} 
    \label{fig:floodsmont}
    \end{minipage}
\end{figure}

From January 1st to December 31st, 2019, there were 280 points of flooding with interdiction of roads in our study area, of which 48\% started in the time slot from 4 pm to 7 pm (Figure \ref{fig:floodsstarttime}). Considering this time slot, about 10.336 million people traveled in the Metropolitan Region of São Paulo (RMSP), of which 4,963,926 were leaving work; and 3,787,260 people were leaving or heading to an educational institution, according to OD Database \cite{METROSP2017}.

The records from December to March represent 70\% of the total (Figure \ref{fig:floodsmont}). That is, the seasonality of the rains is correlated with the occurrences of flooding. Concerning the occurrences, 28\% of floods lasted up to 30 minutes, 23\% from 30min to 1h, 23\% from 1h to 2h; 10\% from 2h to 3h; that is 84\% of the floods lasted up to 3 hours.

Regarding the road hierarchy, 64\% of flooding occurred in arterial roads, 21\% occurred in Collectors roads, 9\% occurred in local roads; and 5\% in Fast Lane (Table \ref{tab:RoadHierarchy}). Roads with the highest frequency of flooding are: Rio Branco (10\%), Mércurio (10\%), Prof. Abraão de Morais (9\%), Prof. Luiz Ignacio Anhaia Mello (7\%), do Estado (6\%), da Bandeira Square (4\%). Except for da Bandeira Square, the other roads are all arterial in the road hierarchy (Table \ref{tab:FloodsbyRoad}). These roads are highlighted in Figure \ref{Fig_Districts}.

\begin{table}[!htb]\centering
\begin{tabular}{r|r|r}
{\textbf{Road Hierarchy}} & {\textbf{Number of floods}} & {\textbf{Floods \%}} \\ \hline
\textbf{Arterial}   & 180 & 64\%   \\
\textbf {Collector}   & 60 & 21\%   \\
\textbf{Local}   & 25 & 9\% \\
\textbf{Fast Lane}   & 14 & 5\% \\
\textbf{Pedestrian Exclusive}   & 1 & 0.4\% \\
\textbf{Total}   & 280 & 100\% \\
\end{tabular}
\caption{Number of floods by road hierarchy.}
\label{tab:RoadHierarchy}
\end{table}

\vspace*{0.1cm}

\begin{table} [!htb]
\centering
\begin{tabular}{r|r|r|r}
{\textbf{Hierarchy}} & {\textbf{Road name}} & {\textbf{Number of floods}} & {\textbf{Floods \%}} \\ \hline
Arterial & Rio Branco Ave & 29 & 10.4 \\
Arterial & Mercurio Ave & 28 & 10.0 \\
Arterial & Prof Abraao de Morais Ave & 24 & 8.6 \\
Arterial & Prof Luiz Ignacio A. Mello Ave & 19 & 6.8 \\
Arterial & do Estado Ave & 16 & 5.7 \\ 
Collector & da Bandeira Square & 12 & 4.3 \\
\end{tabular}
\caption{Number of floods by road (Top 6 roads in flood frequency).}
\label{tab:FloodsbyRoad}
\end{table}

Crossing the location of the flood points with watercourses, it is observed that the flood points are, on average, 300 meters from a watercourse. 50\% of these points are up to 124 meters away from a watercourse. When crossing the location of the flood points with the HAND, the average value is 10.55 m, and 74\% of points are up to 10 m. The crossing of the points with the watercourses and the HAND indicates the presence of two types of flooding: SWF and river flood.

\begin{figure}
    \centering
    \includegraphics[scale=0.7]{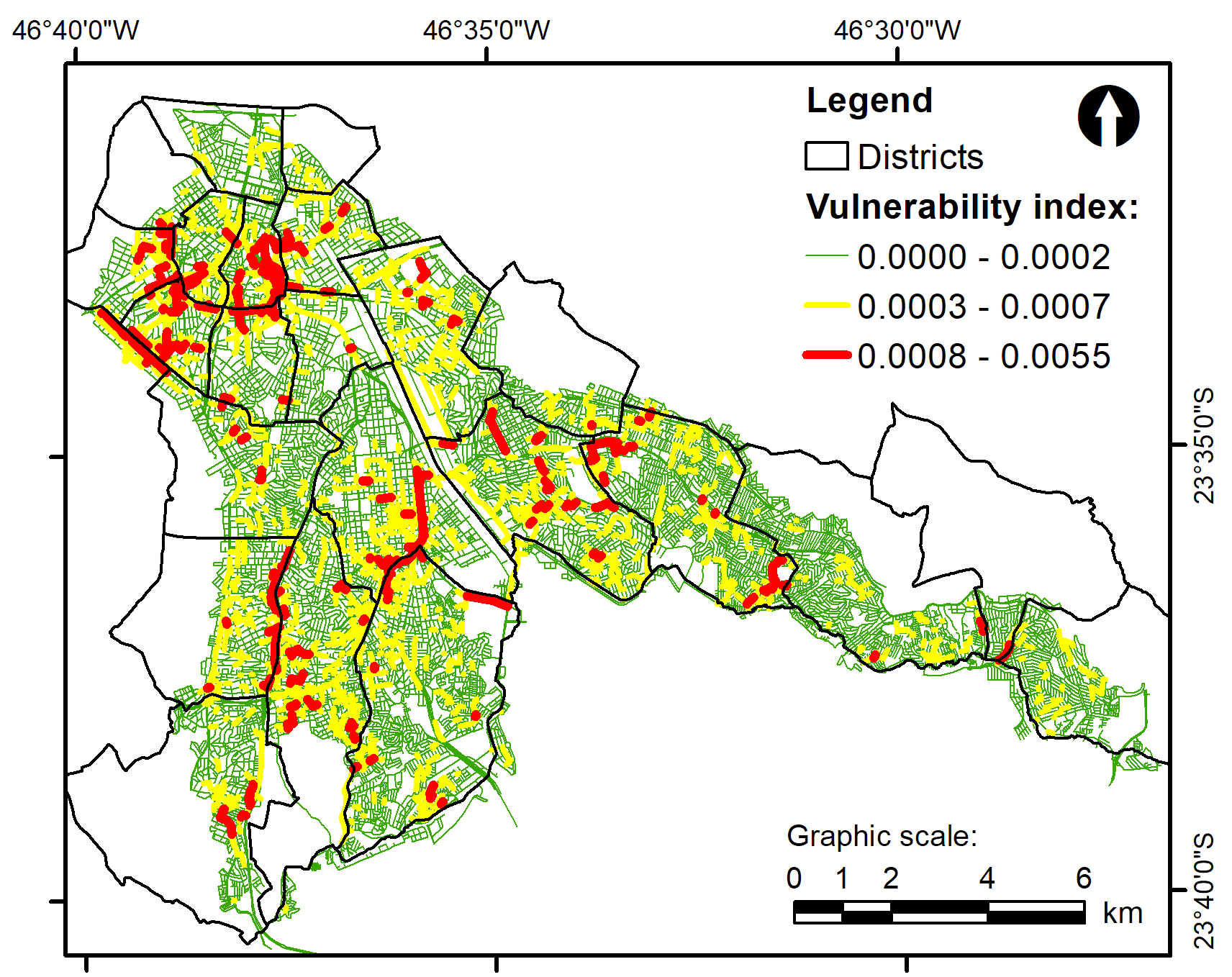} \\
    \caption{Roads vulnerability index.} 
    \label{fig:roadsVulnerab}
\end{figure}

The network is composed of $23,702$ nodes. If one road gives access to another, there is an edge between them. There are $52,921$ edges in the network. The most significant topological distance (number of edges) between one road to another is 39 edges - the network's diameter. On average, the distance between roads is 7.22 intersections, and a road gives access to 4.46 others. The figure \ref{fig:roadsVulnerab} shows the road's vulnerability index, where lower values lead to a less relevant node to the network. The least relevant nodes are those that, when removed, have less impact on the collective characteristics of the network. All vulnerability data (code, inputs, and outputs) is available on GitHub repository \cite{Github}.

\subsection{Analysis of components }\label{section:AllComponents}

Table \ref{tab:Results} summarizes the component results by cells. Except for $FS$, all components have few cells in the Very High class (minimum value of 0.5\% for $LV$, and maximum value of 2.3\% for $NV$). The results are also presented in maps and graphs. The maps allow us to visualize the geographic location of the classes and their neighborhood relationship. The graphs, in turn, allow us to observe the ranges used in the division of the classes of each component (x-axis) and relate them to the frequency (y-axis). The colored divisions of graphs are compatible with the respective map.

\begin{table}[!htb]\centering
\begin{tabular}{r|r|r|r}
{\textbf{Components}} & {\textbf{Moderate}} & {\textbf{High}} & {\textbf{Very High}} \\ \hline
\textbf{$E$} &	2,770 &	288 &	29 \\
\textbf{$LV$} &	3,006 &	65	& 16 \\
\textbf{$NV$}	& 2,512	& 504 &	71 \\
\textbf{$V$} &	2,988	& 82 &	17 \\
\textbf{$PI$} &	3,020 &	58 &	9 \\
\textbf{$FS$} &	575 &	1,027 &	1,485 \\
\textbf{$R$} &	1,583 &	1,467 & 37 \\
\end{tabular}
\caption{Number of cells per Component and Class.}
\label{tab:Results}
\end{table}

\begin{figure} [htb!]
    \centering
    \begin{minipage}{0.5\textwidth}
    \centering
    \includegraphics[scale=0.46]{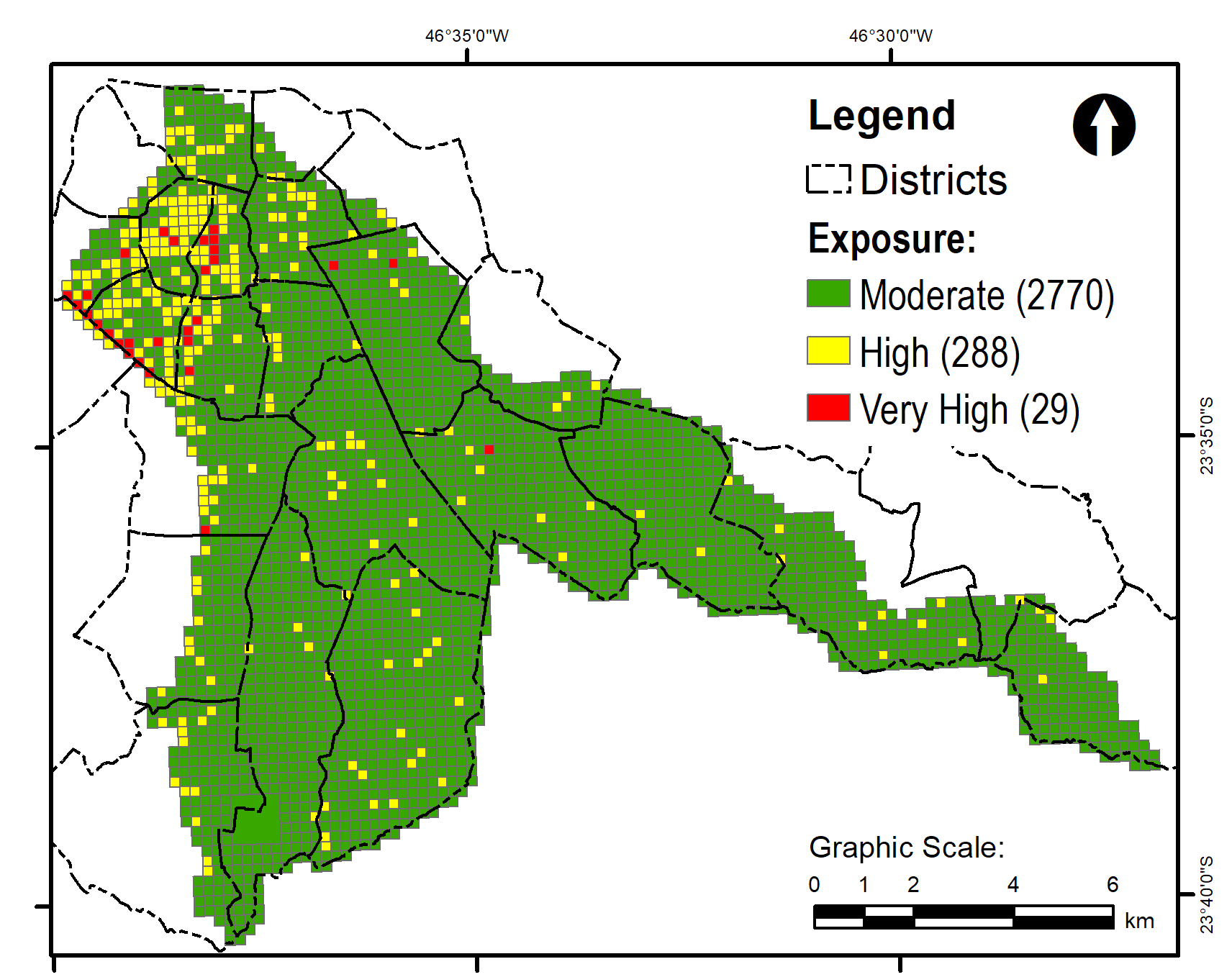} \\
    \caption{Exposure Map.} 
    \label{fig:exposure}
    \end{minipage}\hfill
    \begin{minipage}{0.5\textwidth}
    \centering
    \includegraphics[scale=0.18]{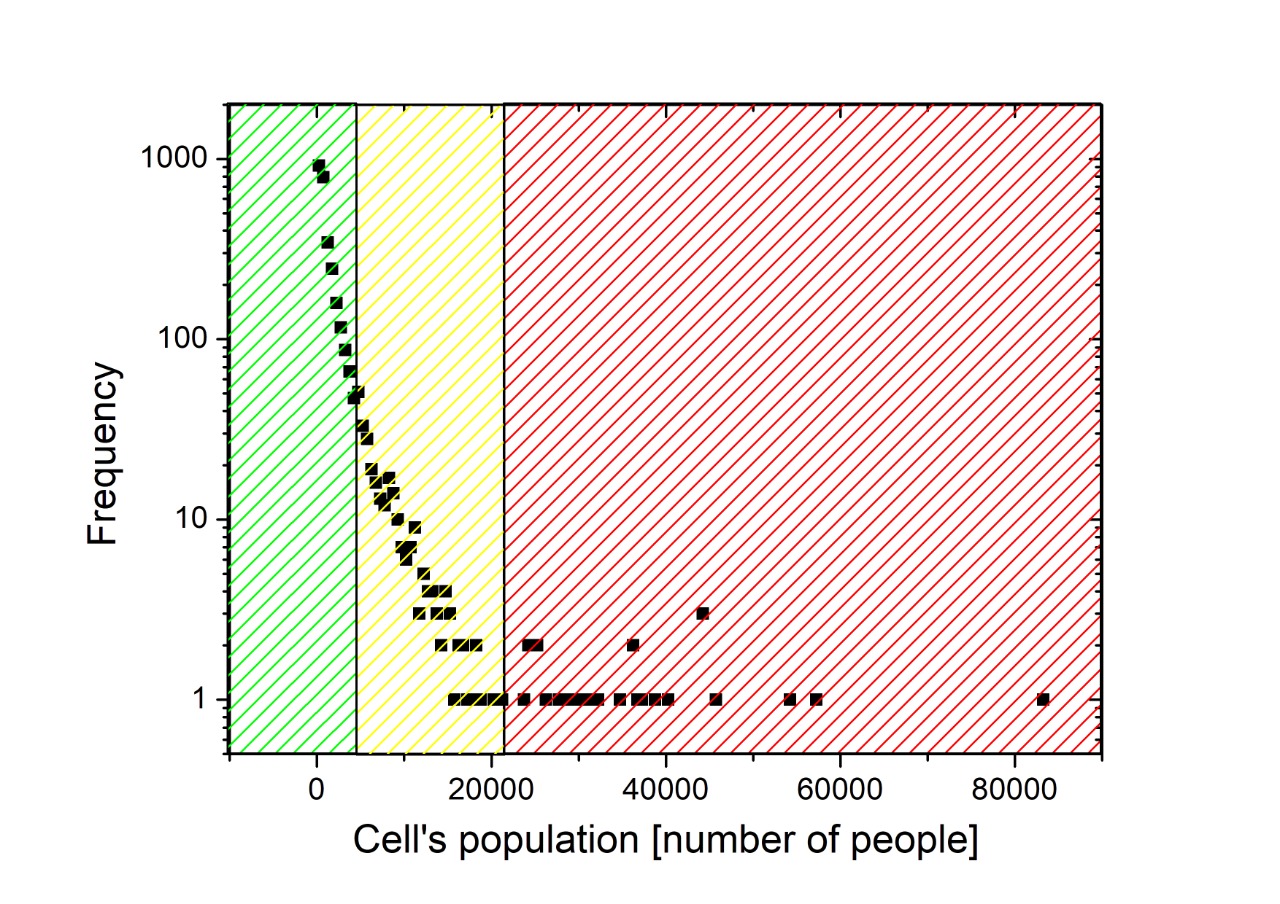} \\
    \caption{Population distribution plot.} 
    \label{graf:population}
    \end{minipage}
\end{figure}

The $E$ component sums 6.4 million people, of which 48\% of people is in the Moderate class; 36\% is in the High class, and 16\% is in the Very High class. Observing the distribution of cells, 0.9\% of the cells in the Very High class and 9.3\% in the High class. These cells are mainly in the northwest portion of the study area, mainly in the historical center and surroundings (Figure \ref{fig:exposure}). It is possible to notice that cells with fewer people are more frequent and are in the Moderate class; and that the greater the number of people per cell, the lower the frequency (Figure \ref{graf:population}).

\begin{figure}[htb!]
    \centering
    \begin{minipage}{0.5\textwidth}
    \centering
    \includegraphics[scale=0.46]{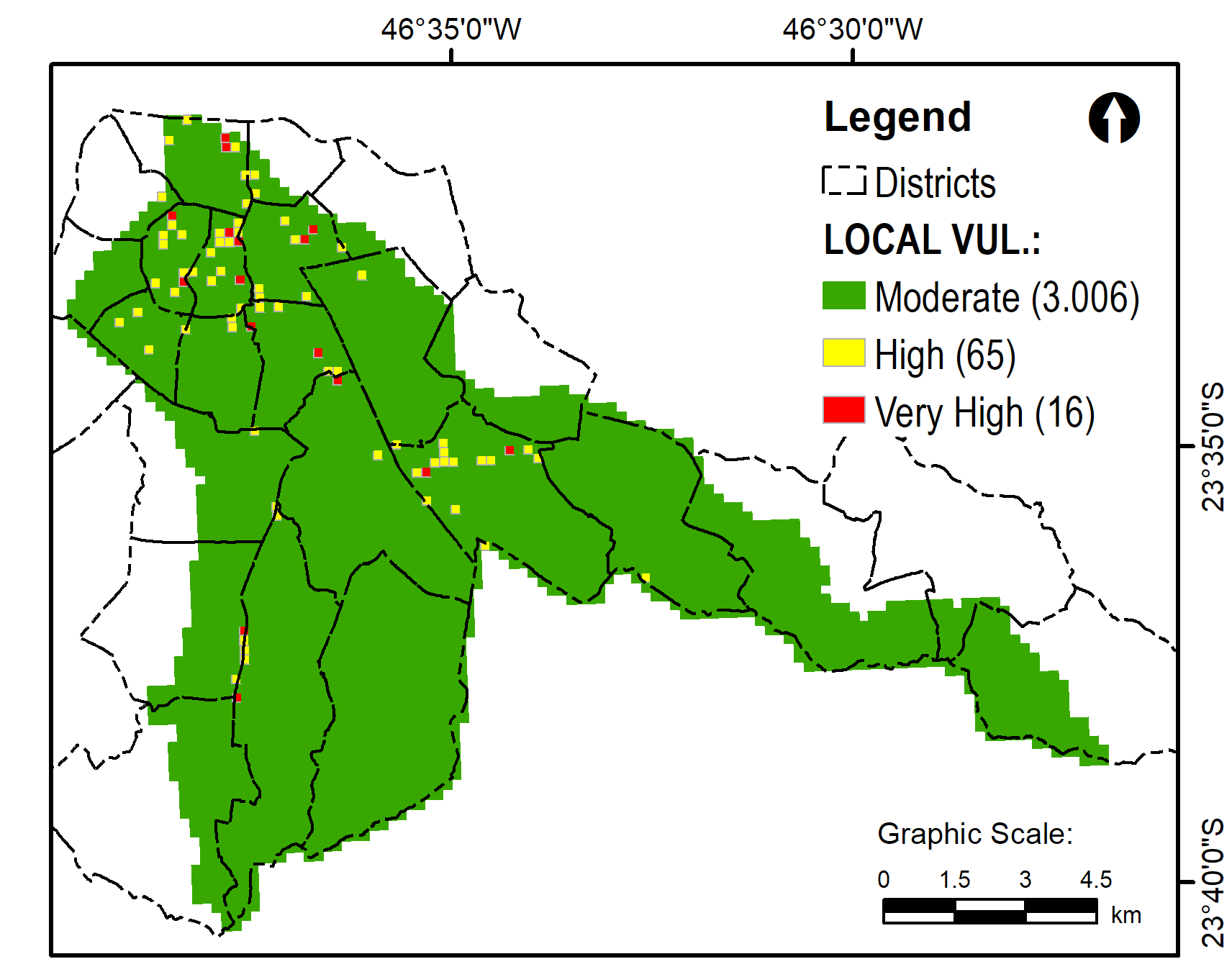} \\
    \caption{Local vulnerability map.} 
    \label{fig:localvulnerability}
    \end{minipage}\hfill
    \begin{minipage}{0.5\textwidth}
    \centering
    \includegraphics[scale=0.18]{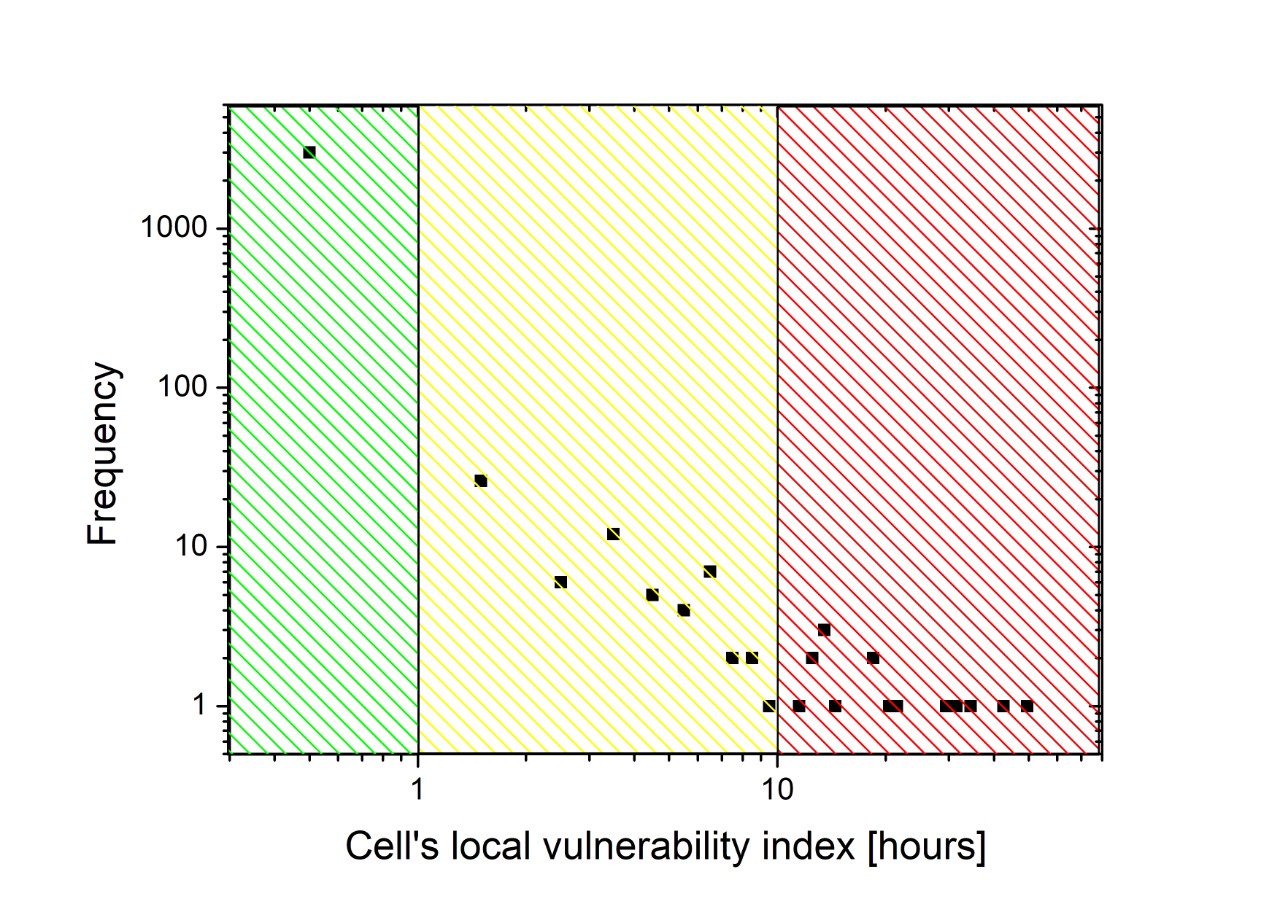} \\
    \caption{Local vulnerability distribution plot.} 
    \label{graf:localvulnerability}
    \end{minipage}
\end{figure}

$LV$ has 0.5\% of the cells in the Very High class and 2.1\% of cells in the High class. These cells are spatially dispersed, but most of them are located in the center of the city (Figure \ref{fig:localvulnerability}). Floods with a shorter duration (up to 1 hour) are more frequent, occupying the Moderate class. Flooding lasting between 1 hour and 10 hours is in the High class, and those that last longer than 10 hours are in the Very High class. The longer the duration, the lower the frequency \ref{graf:localvulnerability}.

The 280 flood points are located in 116 of the 3,087 grid cells. When we consider the duration of these events, we have 3,006 cells with a value of less than one hour. The $LV$ map reflects this characteristic. However, it is essential to highlight that the non-occurrence of flooding in a given place does not mean the absence of future occurrences at this location.

\begin{figure}[htb!]
    \centering
    \begin{minipage}{0.5\textwidth}
    \centering
    \includegraphics[scale=0.46]{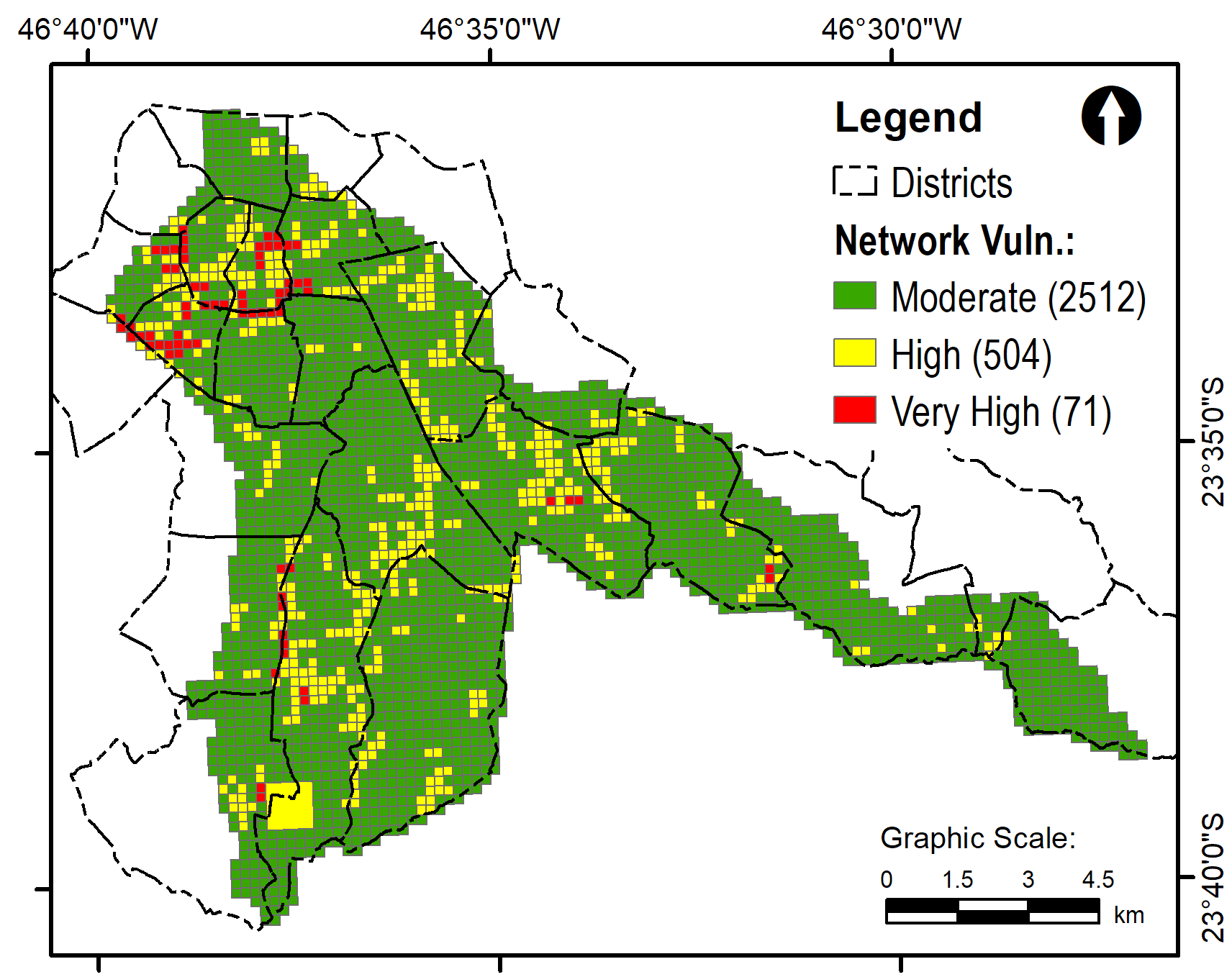} \\
    \caption{Network vulnerability map.} 
    \label{fig:networkvulnerability}
    \end{minipage}\hfill
    \begin{minipage}{0.5\textwidth}
    \centering
    \includegraphics[scale=0.18]{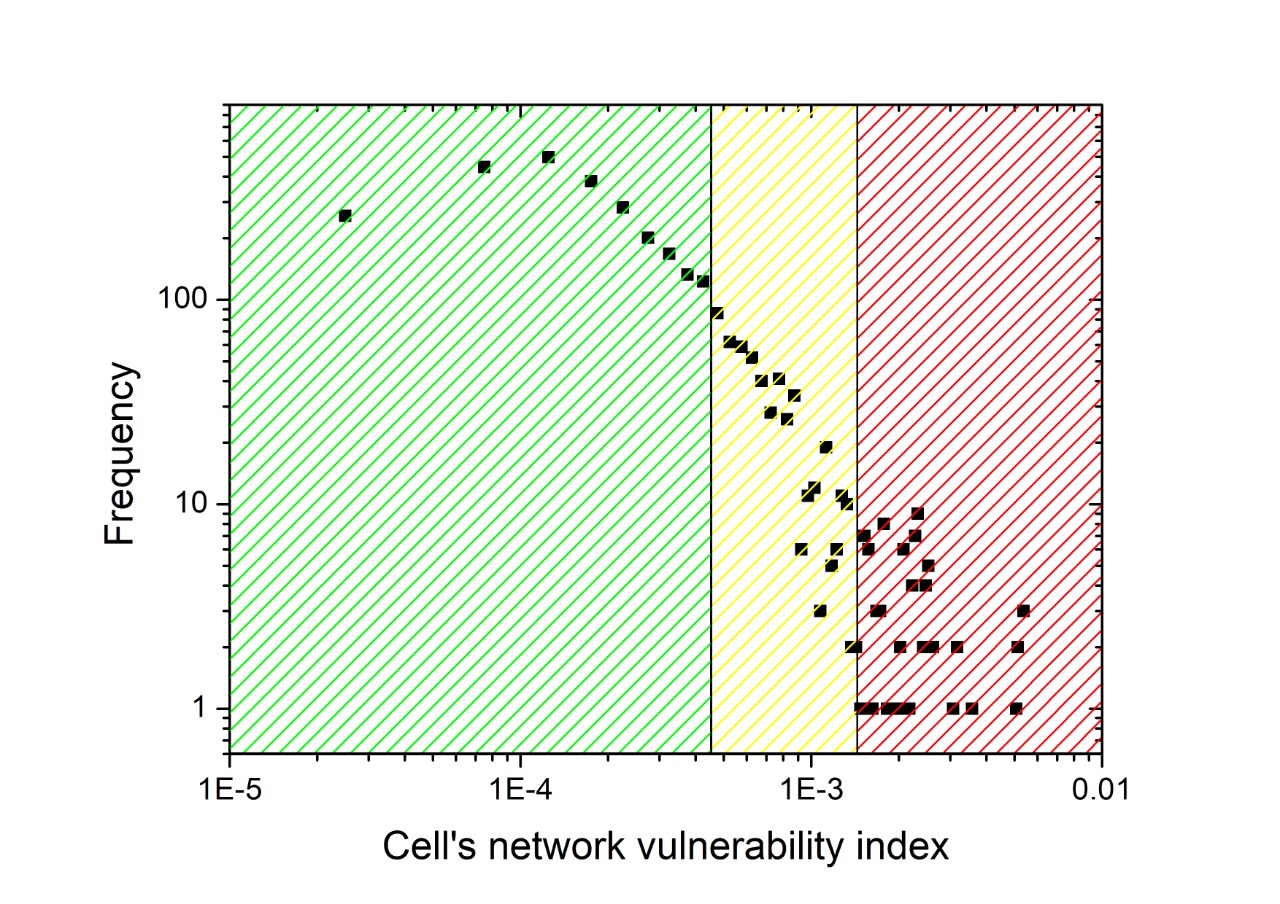} \\
    \caption{Network vulnerability distribution plot.} 
    \label{graf:networkvulnerability}
    \end{minipage}
\end{figure}

$NV$ has 2.3\% of the cells in the Very High class and 16.3\% of cells in the High class. The location of these cells reflects the topology of the input data, and we can relate them to the road axes (Figure \ref{fig:networkvulnerability}). As in the previous graphics, lower values have a greater frequency \ref{graf:networkvulnerability}.

\begin{figure}[htb!]
    \centering
    \begin{minipage}{0.5\textwidth}
    \centering
    \includegraphics[scale=0.47]{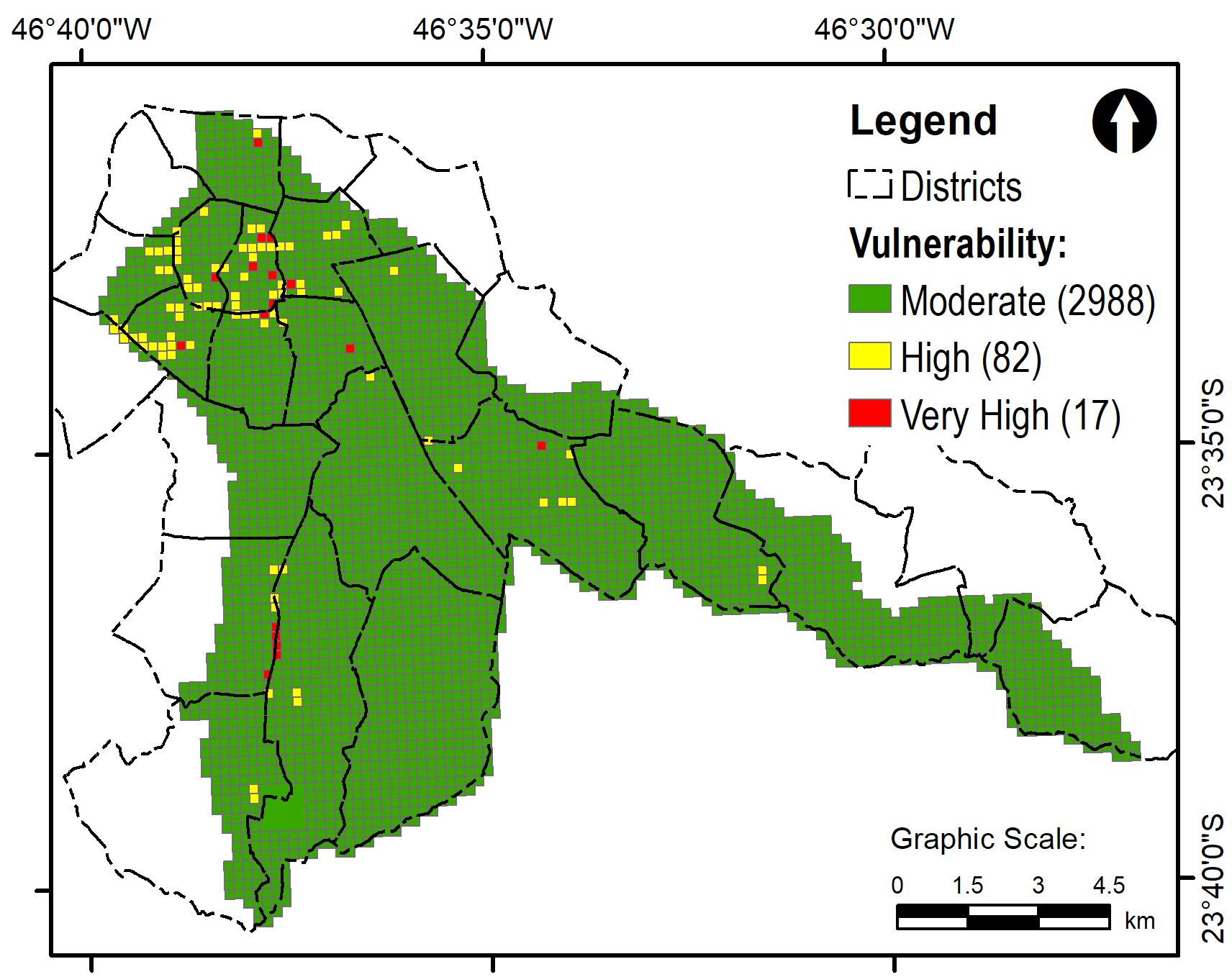} \\
    \caption{Vulnerability map.} 
    \label{fig:vulnerability}
    \end{minipage}\hfill
    \begin{minipage}{0.5\textwidth}
    \centering
   \includegraphics[scale=0.47]{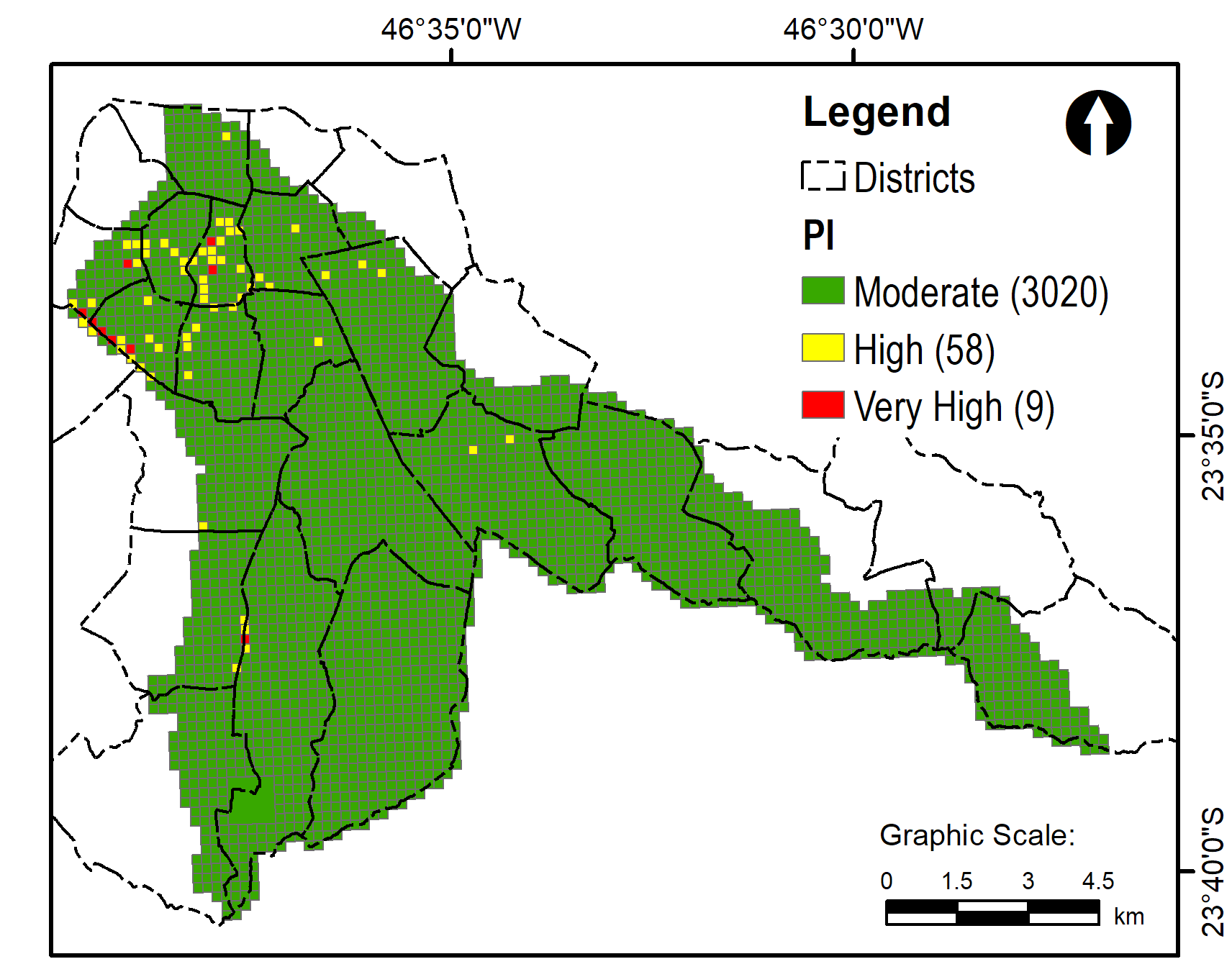} \\
    \caption{Potential impact map.} 
    \label{fig:PI}
    \end{minipage}
\end{figure}

The $V$ has only 0.6\% of the cells in the Very High class and 2.7\% of cells in the High class, with higher concentration in the center (Figure\ref{fig:vulnerability}). It is still possible to perceive some road axes. However, the recurrent location of the flooding causes the scatter of these cells. The integration of the $E$ and $V$ components forms the $PI$, which has 0.3\% of the cells in the Very High class and 1.9\% of cells in the High class located in the same portion of the $V$, but with a smaller number of cells (Figure \ref{fig:PI}).

\begin{figure}[htb!]
    \centering
    \begin{minipage}{0.47\textwidth}
    \centering
    \includegraphics[scale=0.46]{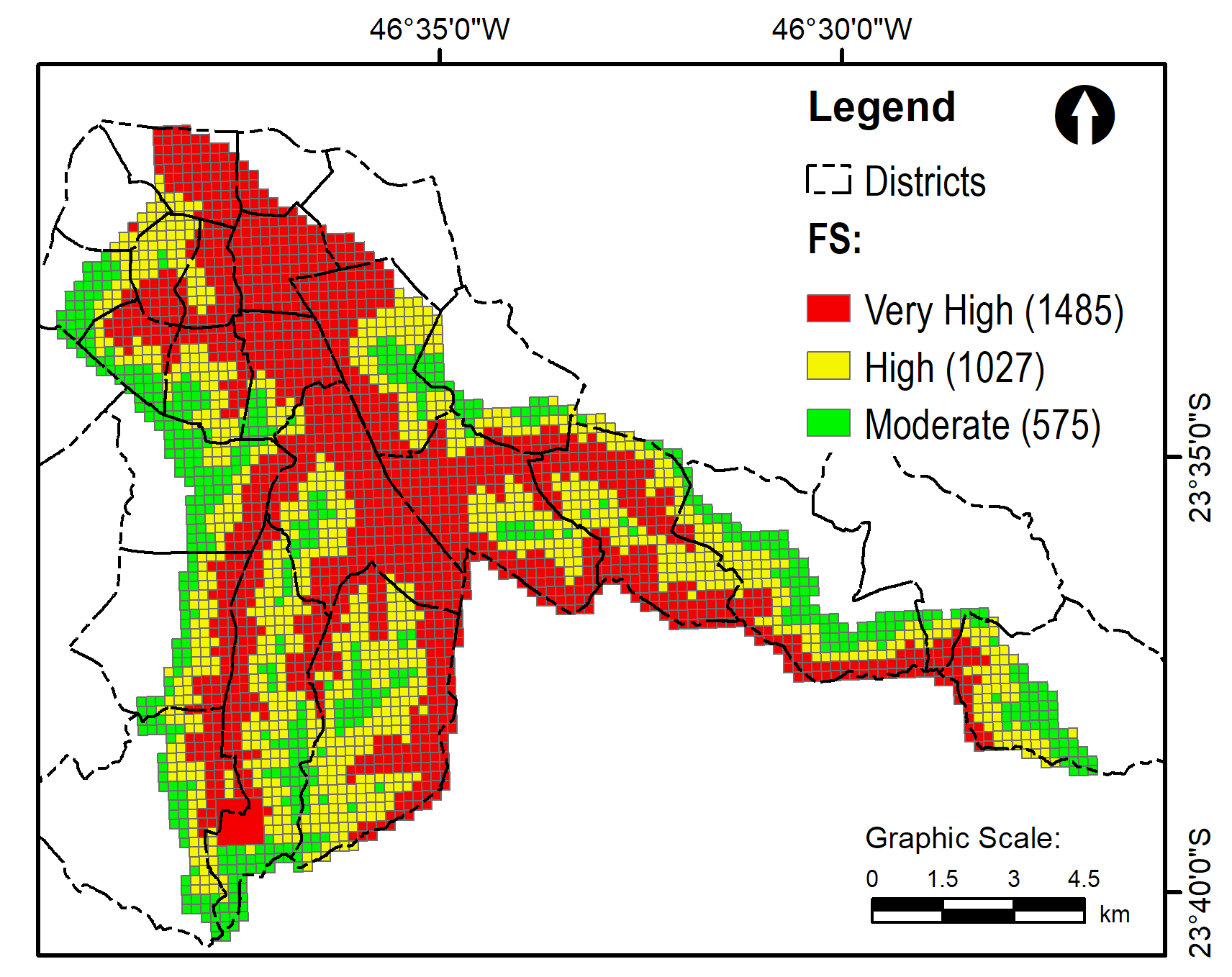} \\
    \caption{Flood susceptibility map.} 
    \label{fig:FS}
    \end{minipage}\hfill
    \begin{minipage}{0.47\textwidth}
    \centering
    \includegraphics[scale=0.18]{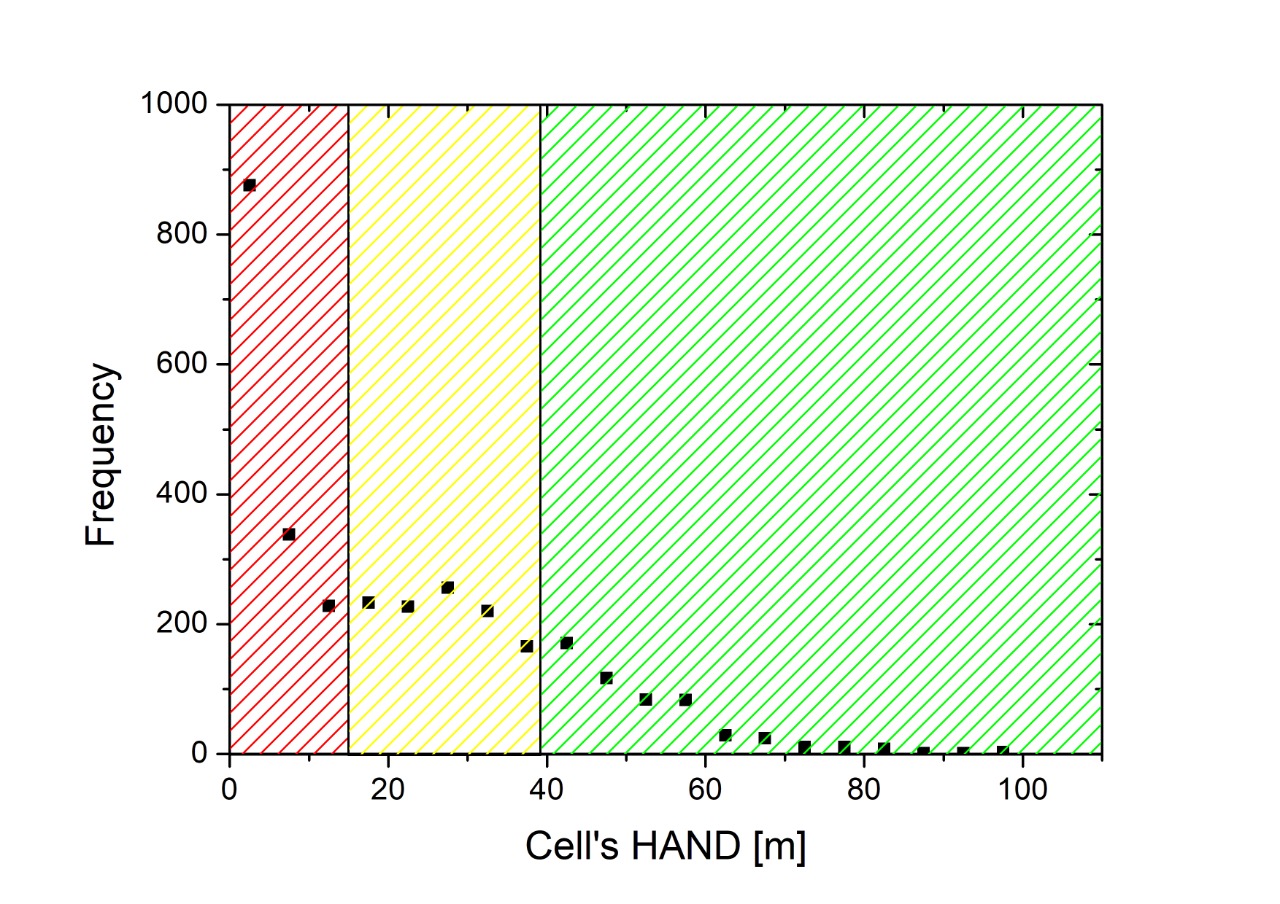} \\
    \caption{HAND distribution plot.} 
    \label{graf:HAND}
    \end{minipage}
\end{figure}

The $FS$ component behaves opposite to the others, with a more significant number of cells (48.1\%) in the Very High class and 33.3\% of cells in the High class, summing up 81.4\% together. This high number in these two classes demonstrates the proximity of cells to watercourses (Figure \ref{fig:FS}). Values up to 15 m are in the Very High class; values between 15 m and 38 m are in the High class, and values above 38 m are in the Moderate class. It means that the lower the HAND, the more significant the $FS$ (Figure \ref{graf:HAND}).

\begin{figure}
    \centering
    \includegraphics[scale=0.5]{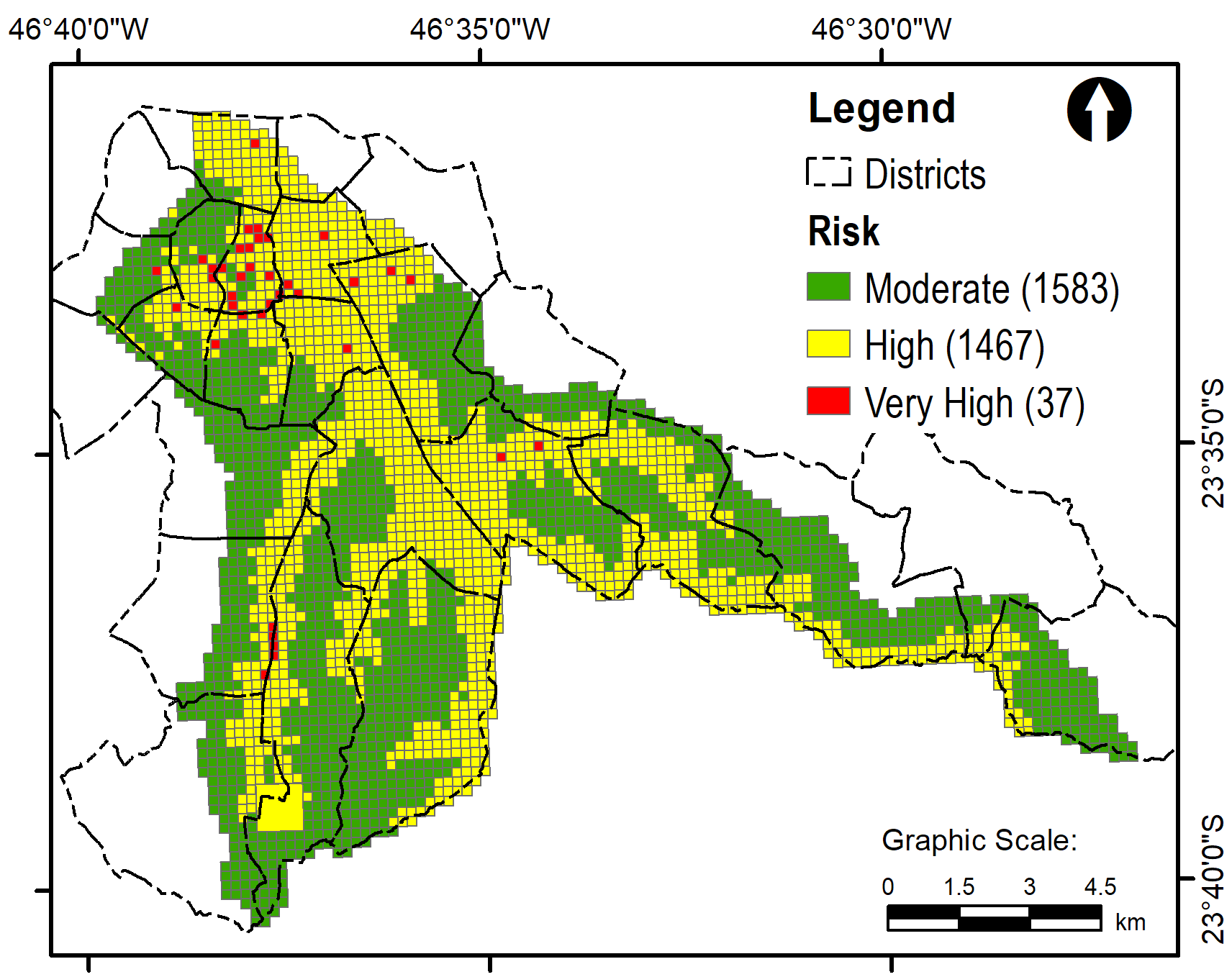} \\
    \caption{Flood risk map.} 
    \label{fig:risk}
\end{figure}

\begin{figure} [hbt!]
    \centering
    \includegraphics[scale=0.36]{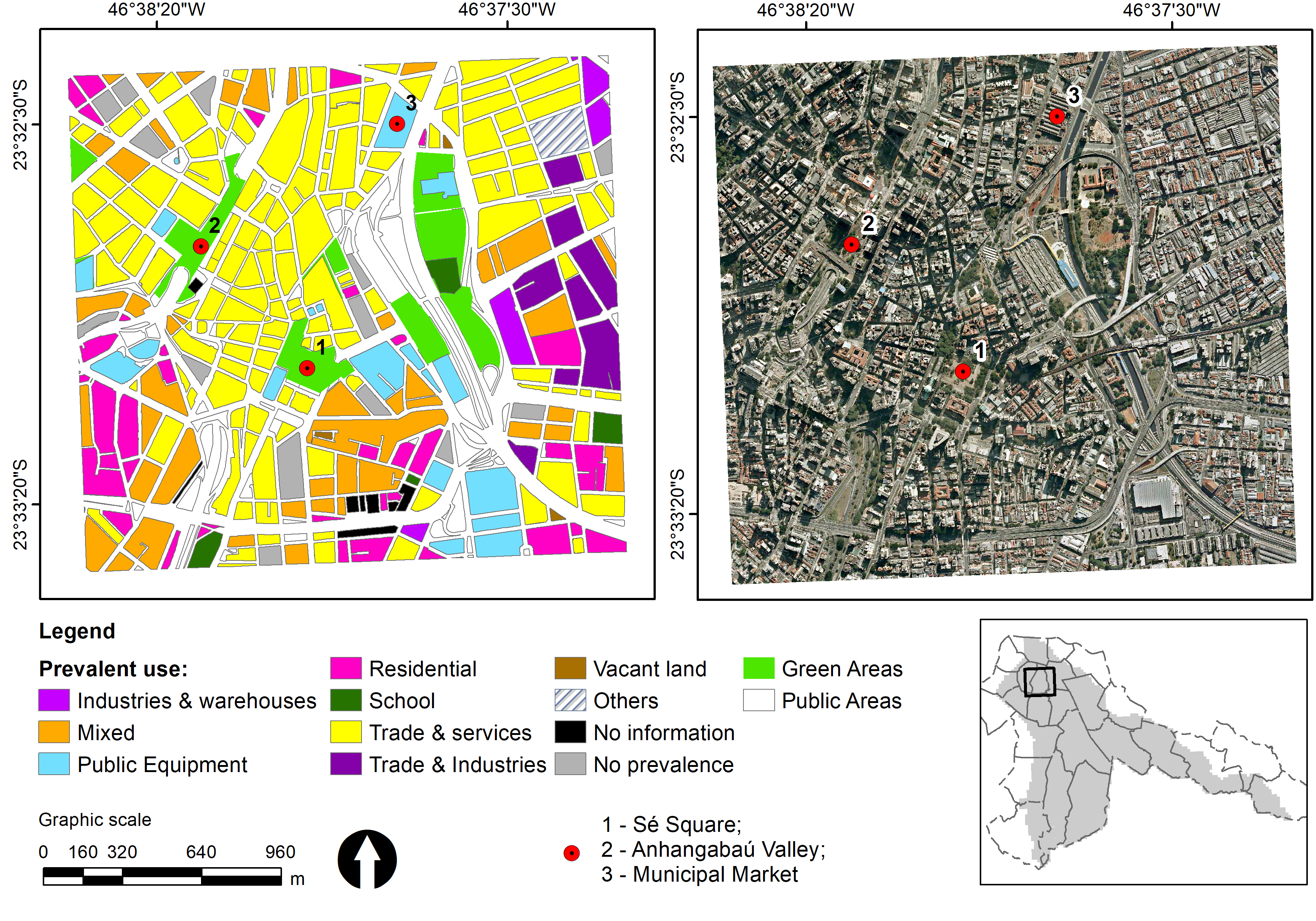} \\
    \caption{Area with the highest concentration of cells in the Very High class.} \label{fig:zoom}
\end{figure}

Finally, the $R$ has 1.2\% of the cells in the Very High class and 47.5\% of cells in the High class (Figure \ref{fig:risk}). 21 of the 37 cells belonging to the Very High class are located in the historic center of the city, covering the Sé Square, the Anhangabau valley, and the Municipal Market  (Figure \ref{fig:zoom}). Although $FS$ strongly influences $R$,  $R$ has only 37 cells in the Very High class. Even if the $FS$ is Very High, the $R$ will only be classified as Very High if the $PI$ is at least classified as High. Therefore, the two components have equal importance in $R$ estimation.

The Secretariat of Urban Development of São Paulo established an aggregation methodology resulting from the crossing between the ``use'' and ``pattern'' values for each registered property generating 16 types of land use. The classification is made by fiscal block, considering the predominant land use (greater than or equal to 60\%) of the properties that make up the block \cite{Geosampa}.

Figure \ref{fig:zoom} shows the land use typology (left) and satellite image (right) of the central area with the highest concentration of cells classified as Very High risk. Analyzing this clipping, it is possible to notice that it is a built-up area with impervious surfaces with a high density of commercial buildings (50\% of the blocks are classified as commerce and services). This is an area with great attraction for travel because of this predominant typology.

\section{Conclusions and Perspectives}\label{section:Conclusions}

This article proposes an urban flood risk map from a multidisciplinary perspective, integrating data from different sources and geographic units into a regular grid. The methodology allows the use of components in an isolated or integrated way, with several possible components' combinations. In our study, all components, except $FS$, have few cells in the Very High class (minimum value of 0.5\% in the case of $LV$ and maximum value of 2.3\% for $NV$). $FS$ component reflects the presence of watercourses, and it has a strong influence on the location of cells classified as High. The cells classified as Very High risk are primarily in downtown, a built-up area with impervious surfaces.

In addition, the $FS$ can be used to better position teams when it starts to rain; and the potential impact can help prioritize care when there are two simultaneous floods. The flood risk map enables policymakers to figure out where to allocate resources as it is also helpful for emergency managers, the exposed population, land-use planners, and infrastructure owners. They are all potential users of flood maps. Flood maps are a vital tool in flood risk management, including communication. They are more assertive in visualizing the spatial distribution of the flood risk than other forms of presentation, increasing awareness of decision-makers and people at risk. Furthermore, flood maps can serve as a base for deriving flood insurance premiums or allowing disaster managers to prepare for emergencies.

Exploratory analysis helps not only to delineate the methodology but also to understand how the variables behave. This tool allows us to know the parts that make up the urban space and how they are related. In our study, flooding has a seasonal behavior with more expressive numbers in the summer. The exposed population will also vary according to the day and time, especially in areas with a predominance of commerce and services. These variations can be explored in a dynamic risk map, with seasonal or fluctuating risks throughout the day.

Another critical issue is to determine the areal unit of analysis since each variable has a different boundary. We work with punctual (workplace, study places, flooding), linear (road system), and matrix (statistical grid and HAND) data, which are incompatible systems. We chose to aggregate this data into a square grid cell to handle official data and harmonize different spatial data. This topic should receive attention in methodology, as it will always be recurrent in multidisciplinary studies.

The methodology can also be adapted, including new variables, to better reflect the dynamics of complex systems. We made an advance in estimating the population exposed to floods by including the population that works or studies in the analyzed area. The aggregation of the population took place in the final destinations of the trip (workplaces and study institutions) without considering the route taken.

As a future investigation, we suggest including dynamic data, which considers how the population moves until reaching the final destination, whether by volumetric vehicle counting, crowdsourced data, or specific mobility surveys. And, to improve the calculation of the network vulnerability index, we are developing, as future work, an index by event to reach more than one element at the same time.



\section*{ACKNOWLEDGEMENTS}

We thank Professor Carlos Augusto Morales Rodriguez for the valuable initial discussions about flood data in Sao Paulo. Also, we thank the Intelligent Systems Computing Laboratory (CSILab) from the Federal University of Ouro Preto for sharing computational resources and the financing institutions that supported us in the development of this research.

\section*{CONFLICT OF INTEREST}
No potential conflict of interest was reported by the author(s).



\printendnotes

\bibliography{2_Bibliog}




\end{document}